\documentclass{article}

\usepackage{arxiv}

\usepackage[utf8]{inputenc} 
\usepackage[T1]{fontenc}    
\usepackage{hyperref}       
\usepackage{url}            
\usepackage{booktabs}       
\usepackage{amsfonts}       
\usepackage{nicefrac}       
\usepackage{microtype}      
\usepackage{lipsum}		
\usepackage{graphicx}
\usepackage{natbib}
\usepackage{doi}

\usepackage{comment}
\usepackage{amsmath}
\usepackage{cleveref}

\title{Complementarity in Human-AI Collaboration: \\ Concept, Sources, and Evidence}

\author{ 
  Patrick Hemmer \\
  Karlsruhe Institute of Technology \\
  \texttt{patrick.hemmer@kit.edu} \\
  \And
  Max Schemmer \\
  Karlsruhe Institute of Technology \\
  \texttt{max.schemmer@kit.edu} \\
  \And
  Niklas Kühl \\
  University of Bayreuth \\
  \texttt{kuehl@uni-bayreuth.de} \\
  \And
  Michael Vössing \\
  Karlsruhe Institute of Technology \\
  \texttt{michael.voessing@kit.edu} 
  \And
  Gerhard Satzger \\
  Karlsruhe Institute of Technology \\
  \texttt{gerhard.satzger@kit.edu} \\
}

\renewcommand{\shorttitle}{Complementarity in Human-AI Collaboration: Concept, Sources, and Evidence}

\hypersetup{
pdftitle={Complementarity in Human-AI Collaboration: Concept, Sources, and Evidence},
pdfsubject={Human-AI Complementarity, Human-AI Decision-Making, Complementary
Team Performance, Human-AI Teams, Human-AI Collaboration, Complementarity Potential},
pdfauthor={Patrick Hemmer, Max Schemmer, Niklas Kühl, Michael Vössing, Gerhard Satzger},
pdfkeywords={Human-AI Complementarity, Human-AI Decision-Making, Complementary
Team Performance, Human-AI Teams, Human-AI Collaboration, Complementarity Potential}, 
}

\begin{document}
\maketitle

\begin{abstract}
Artificial intelligence (AI) has the potential to significantly enhance human performance across various domains. Ideally, collaboration between humans and AI should result in complementary team performance (CTP)---a level of performance that neither of them can attain individually. So far, however, CTP has rarely been observed, suggesting an insufficient understanding of the principle and the application of complementarity. Therefore, we develop a general concept of complementarity and formalize its theoretical potential as well as the actual realized effect in decision-making situations. Moreover, we identify information and capability asymmetry as the two key sources of complementarity. Finally, we illustrate the impact of each source on complementarity potential and effect in two empirical studies. Our work provides researchers with a comprehensive theoretical foundation of human-AI complementarity in decision-making and demonstrates that leveraging these sources constitutes a viable pathway towards designing effective human-AI collaboration, i.e., the realization of CTP.
\end{abstract}


\section{Introduction}

The increasing capabilities of artificial intelligence (AI) have paved the way for collaborating with humans and supporting them in a wide range of domains. Examples include decision support for humans in application areas such as customer services \citep{Vassilakopoulou2023}, medicine \citep{Jussupow2021}, law \citep{Mallari2020}, finance \citep{Day2018}, and industry \citep{Stauder2021}. With AI decisions becoming increasingly accurate, there is an obvious temptation to fully rely on them and to automate decision tasks. However, this approach often falls short of realizing even better performance by combining and integrating the unique strengths of the individual members in a human-AI team \citep{Seeber2020}. The recent emergence of large language models illustrates this \citep{Malone2023}: While applications like ChatGPT often provide helpful, but not always correct results, a human decision-maker can collaborate with the system to, for example, override erroneous responses in order to achieve superior task performance \citep{Malone2023}. Similarly, in the medical domain, both AI models and physicians are able to produce diagnoses individually. It has, however, been demonstrated that humans and AI models could make different errors \citep{Geirhos2021,Steyvers2022} so that they may realize superior results when ``teamed up'': For instance, the AI model might detect patterns in large amounts of data that humans might not discover easily, while humans might excel at the causal interpretation and intuition required to understand these patterns \citep{Lake2017}.

This complementarity as the ``quality of being different but useful when combined'' \citep{Cambridge_Dictionary} has inspired researchers to investigate how humans’ and AI’s individual abilities could be leveraged to achieve superior team performance compared to either one performing the decision task independently. Such an outcome is defined as \textit{complementary team performance (CTP)} \citep{Bansal2021}. This phenomenon is of increasing interest as recent studies tell only part of the story: On the one hand, various studies have demonstrated that human-AI teams are able to outperform human individuals \citep{Alufaisan2021,Inkpen2023,Liu2021,Sarkar2023}---often not analyzing, though, whether they also surpass the AI model's individual performance \citep{Bansal2021}. On the other hand, while many settings exist in which humans still show better task performance \citep{Grace2018,Grace2024}, recent studies also demonstrate evidence for human-AI teams outperforming AI models \citep{Dvijotham2023,Fugener2021a,Ma2023}.

To make things even more complex, the particular design of human-AI collaboration in decision-making affects humans and their contributions within the team, e.g., the use of unique human knowledge \citep{Fugener2021}, the self-assessment of human capabilities \citep{Fugener2021a}, the adjustments of mental models \citep{Bauer2023a}, the incentive for ``active rethinking'' \citep{Lu2024}, or human task performance and learning \citep{Foerster}. In summary, current research still misses compelling explanations for the success of human-AI teams as well as a systematic understanding of complementarity when the team performance is measured against the performance of both team members individually. Thus, concentrating on decision-making as an important application of human-AI collaboration \citep{Lai2021}, we pursue the following two research questions in this work:

\textbf{RQ1:} How can we model human-AI collaboration in decision-making to enable a more nuanced understanding of the synergetic potential in a human-AI team?

\textbf{RQ2:} What factors contribute to complementary team performance?

We address these research questions by developing a conceptualization of human-AI complementarity that introduces and quantifies \textit{complementarity potential (\(CP\))} and \textit{complementarity effect (\(CE\))} and outlines the two key sources of complementarity—information and capability asymmetry. In detail, we argue that both complementarity potential and complementarity effect consist of an inherent and a collaborative component. Whereas the first component captures decision-making synergies that, for each task instance, can be attributed to the individually more accurate team member within the human-AI team, the second component captures decision-making synergies that only emerge through collaboration resulting in team decisions which are more accurate than each of the individual ones. We demonstrate the application and the value of our conceptualization in two experimental studies—leveraging the two sources of complementarity, i.e., information and capability asymmetry within the human-AI team. In both studies, humans collaborate with an AI model to conduct decision-making tasks. The AI model provides independent decision suggestions that humans can incorporate into their judgment to derive a final team decision. In the first study, we choose the domain of real estate valuation to investigate information asymmetry. We train an AI model to predict real estate prices based on tabular data. Humans receive its suggestions and also have access to a photograph of the real estate. They can use both to arrive at a final team decision. In the second study, we select the context of image classification to analyze the impact of capability asymmetry between humans and AI. We train two AI models whose capability gaps differ from those of the human decision-maker. In both studies, we apply our conceptualization to demonstrate that both of these sources increase the inherent component of complementarity potential as well as the realized effect, resulting in CTP.

To summarize, we make the following contributions to the state of knowledge in information systems (IS) research: First, we conceptualize human-AI complementarity as a means to comprehensively analyze and design human-AI collaboration in decision-making. Second, we scrutinize information and capability asymmetries as sources of complementarity. Third, we demonstrate the value and application of our concepts and the sources’ potential impact in two behavioral experiments. This should provide IS researchers with a better understanding and methodological support when purposefully designing human-AI collaboration in decision-making for more effective outcomes---thereby supporting the development of hybrid intelligence \citep{Dellermann2019} and advocating the ``AI \textit{with} human'' (opposed to ``AI \textit{vs.} human'') perspective in societal debates on the future of work \citep{Huysman2020}.

In the remainder of this work, we first outline the relevant background and related work in \Cref{sec:related_work_ejis}. In \Cref{sec:concpetualization_human_ai_complementarity_ejis}, we derive the conceptualization of human-AI complementarity. In \Cref{sec:experimental_studies_ejis}, we empirically illustrate our conceptualization’s utility in two experimental studies. We discuss our results in \Cref{sec:discussion_ejis}, before concluding the work in \Cref{sec:conclusion_ejis}. We provide the Appendix of this paper at \url{https://github.com/ptrckhmmr/human-ai-complementarity}.

\section{Theoretical Foundations and Related Work}\label{sec:related_work_ejis}

In this section, we elaborate on the key concepts and on existing work that provide the foundation for a formal conceptualization of complementarity. We first cover existing differences between humans and AI models\footnote{Throughout this paper, we refer to models or systems using machine learning as ``Artificial Intelligence'' (AI) \citep{Berente2021,Collins2021,Rai2019}. While we acknowledge the technical distinction between AI and Machine Learning (ML) as discussed, e.g., by \citet{Kuhl2022} and although we have considered the more precise reference to an ``ML'' model, we adopt the use of the broader ``AI'' term as the prevalent terminology established in the Computer Science and Human-Computer Interaction communities. This choice reflects the contemporary linguistic trend rather than a lack of distinction between the two fields.} as sources of complementarity, then review existing literature on human-AI collaboration in decision-making as the means to harness complementarity, and finally summarize the current state of knowledge on human-AI complementarity.

\subsection{Sources of Complementarity: Information and Capability Asymmetry}\label{sec:sources_of_complementarity}

\begin{figure}[h]
    \centering
    \includegraphics[width=0.6\textwidth]{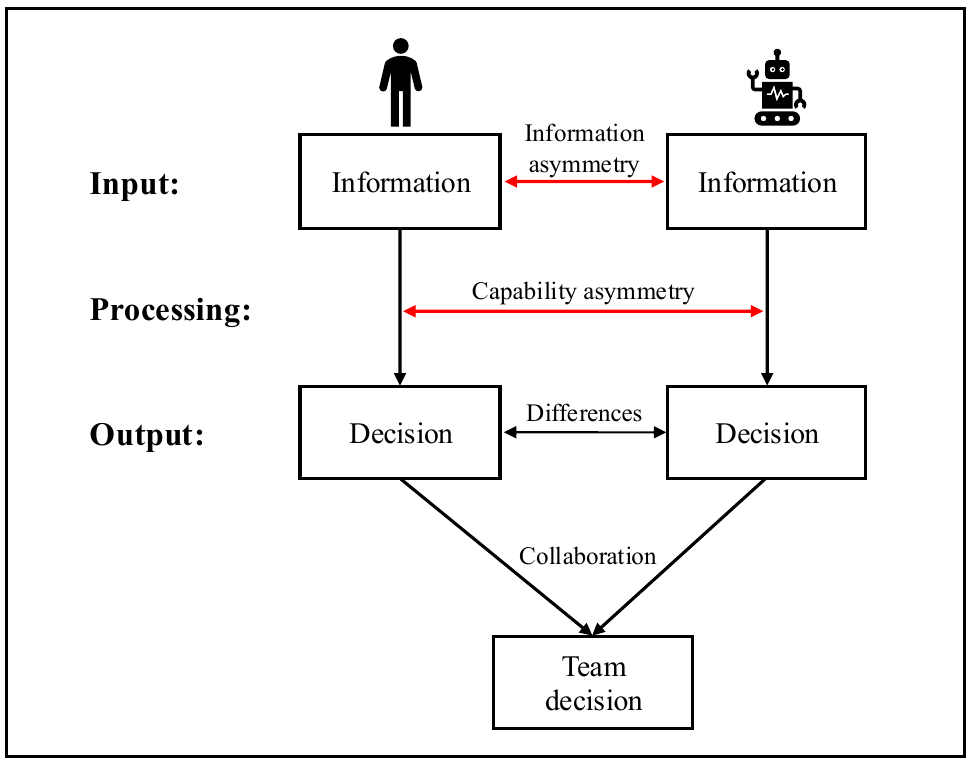}
    \caption{Asymmetries in decision-making between human and AI as sources of complementarity.}
    \label{fig:asymmetries_decision_making}
\end{figure}

In the following, we assume that both human and AI can independently produce solutions to a particular decision problem. Collaboration would not be of interest if both always generated identical decisions. However, human and AI typically make different types of errors \citep{Geirhos2021,Steyvers2022}---a phenomenon that very generally can be traced back to two key sources. To illustrate this, we employ a simple Input-Processing-Output Model for decision-making (\Cref{fig:asymmetries_decision_making})---as used both in IS \citep{Arcy,Parsons2012} and social psychology \citep{Gladstein1984,Hackman1975}: Any team player (human or AI) draws on and processes certain sets of information to derive a decision. Discrepancies in the decision outcome may either stem from different levels of available information (\textit{information asymmetry}) or from different capabilities to process this information (\textit{capability asymmetry}). The two asymmetries may then be exploited for a (hopefully) improved team decision resulting in CTP.

\paragraph{Information asymmetry.} Often, humans and AI have access to different sets of information as decision input: On the one hand, AI models are trained on a well-defined, limited, and digitally available set of data \citep{LeCun2015}. Humans, however, may also have access to information that—due to technical or economic reasons—is not digitized and, thus, not usable for an AI model \citep{Ibrahim2021}. They may use contextual information or information on rare events for a more holistic decision-making setting. For example, AI models in the medical domain derive diagnoses from X-ray scans by analyzing the pixels of the image \citep{Irvin2019}. Human radiologists, however, may also draw on information from direct interaction with patients or from access to medical records. On the other hand, the AI may also have access to information not available to the human decision-maker: A driving assistance system may base its actions on sensor data, e.g., from lidar systems that are not accessible by the human while driving \citep{Li2020}.

\paragraph{Capability asymmetry.} Even if the information would be identical, different outcomes may be driven by different internal modes of information processing between human and AI \citep{Kühl_goutier}. Such capability asymmetries emerge as AI models encode relationships inferred from training data, whereas humans employ compositional mental models that encode beliefs about the physical and social world \citep{Lake2017,Rastogi2023}. While humans are capable of conducting decision-making tasks already after a few trials, AI models require vast amounts of training data to become capable of accurately performing decision-making tasks \citep{Gopnik2012,Kühl_goutier,Lake2017,Tenenbaum2011}. Similarly, humans have often gained experiences with regard to a particular decision-making task continuously over their lifetime, while AI models’ experiences are limited to the task instances seen during model training and may only be updated from time to time \citep{Dellermann2019,Rastogi2023}. On the other hand, AI may excel in digesting vast amounts of information much faster than humans could, and its processing exhibits a greater capability to perceive even small variations in data \citep{Findling2021}.

Overall, both asymmetries may give rise to performance synergies when collaborating in human-AI teams. A traditional example is forecasting theory \citep{Sanders1991,Sanders1995,Sanders2001}: \citet{Sanders1995} analyze the effects of combining statistical forecasts with predictions of a human with access to contextual information and find that such combinations could positively impact forecast accuracy. Composing suitable teams to optimize team performance has been investigated for human-only teams \citep{Horwitz2005}, but lately has also been applied to human-AI teams \citep{Hemmer2022}. While research has shown that such asymmetries can improve team performance \citep{Simons1999,Zhang2022}, they also bear the risk of negative effects, i.e., performance degradation \citep{Dougherty1992}: Heterogeneous and interdisciplinary human-only teams may enjoy benefits from higher levels of individual problem-solving creativity that may be outweighed, though, by the difficulties to effectively communicate with each other \citep{Ancona1992}.

\subsection{Means for Harnessing Complementarity: Human-AI Collaboration}\label{sec:means_for_harnessing_complementarity}

Many terms are used to describe the interplay between humans and AI. Common ones are \textit{human-AI team} \citep{Seeber2020}, \textit{human-AI collaboration} \citep{Vossing2022}, and \textit{human-AI decision-making} \citep{Lai2021}. These are interrelated concepts that emphasize combining the complementary qualities of humans and AI. The notion of a \textit{human-AI team} refers to an organizational setup in which AI systems are increasingly considered equitable team members rather than support tools for humans---since AI can perform a continuously growing number of tasks independently \citep{Endsley2023,Seeber2020}. \textit{Human-AI collaboration} is the process in which these teams work together in a synergistic manner to achieve shared goals, e.g., with the AI providing recommendations or insights, and humans guiding and refining the AI-generated outputs \citep{Terveen1995,Vossing2022}. \textit{Human-AI decision-making} refers specifically to human-AI collaboration in decision-making tasks. For example, the AI could offer data-driven decision recommendations, while humans leverage their domain expertise, emotional intelligence, and ethical considerations to combine AI recommendations with their judgment to reach a final team decision \citep{Lai2021}.

In this article, we focus on human-AI decision-making as a key application area of human-AI collaboration. The increasing abilities of AI have contributed to its use in a growing number of application domains, such as medicine \citep{McKinney2020}, finance \citep{Kleinberg2018}, and customer services \citep{Vassilakopoulou2023}. Consequently, AI-based technologies are employed in processes and systems with varying degrees of human involvement, ranging from autonomous decision-making \citep{Rinta-Kahila2022} to just auxiliary support for the ultimate human decision-maker \citep{Bansal2021,Bucinca2020,Lai2020,Liu2021}. In this context, unintended or unfair outcomes, e.g., AI-based systems’ decisions that benefit certain individuals more than others \citep{Kordzadeh2022}, have ignited a debate on the degree of autonomy granted to AI to ensure responsible outcomes \citep{Mikalef2022}. To alleviate possible detrimental effects, configurations have been suggested that keep humans in the decision-making loop \citep{Gronsund2020,Pathirannehelage,Mikalef2022}.

Researchers have been devoting significant efforts to better understand human-AI decision-making and to design the collaboration in a way that ultimately achieves CTP \citep{Hemmer2021,Lai2021}. Overall, a wide range of behavioral experiments seek to help us understand how humans make decisions within human-AI teams \citep{Alufaisan2021,Bansal2021,Bucinca2020,Carton2020,Fugener2021,Fugener2021a,Lai2020,Liu2021,Malone2023,Reverberi2022,VanderWaa2021,Zhang2022,Zhang2020}. A key emerging concept in this space is that of human reliance on AI advice that needs to be appropriately calibrated to ensure effective decision-making \citep{Bucinca2020,He2023,Kunkel2019,Schemmer2022,Yu2019,Zhang2020}. To assist the human in judging the AI’s decision quality, the AI can provide information about the decision’s uncertainty \citep{Fugener2021,Zhang2020}. Similarly, it may deliver various types of explanations that shed light on its decision-making rationale \citep{Adadi2018,Bauer2023a}, e.g., using feature-based \citep{Ribeiro2016}, example-based \citep{VanderWaa2021} or rule-based techniques \citep{Ribeiro2018}. In application domains involving high-stakes decisions, e.g., medicine, it is crucial for human experts not to exhibit a certain reliance ``on average'' but in particular to identify the AI model’s incorrect suggestions \citep{Jussupow2021}. Thus, the calibration of reliance needs to reflect the actual decision quality \citep{Schoeffer2023}.

A closer look at quantitative studies on human-AI decision-making reveals that, in general, human performance increases when supported by high-performing AI models. In the vast majority of cases, however, the team performance remains inferior to that of the AI model when performing the task alone \citep{Hemmer2021,Malone2023}. This means that joint decision-making currently does not lead to the realization of the full complementarity potential: Humans often do not show appropriate reliance by contributing their own decision capabilities in the right places. While recent studies have shown that the performance of human-AI teams \textit{can} improve beyond that of the individual team members \citep{Dvijotham2023,Fugener2021a,Ma2023}, the underlying mechanisms \textit{why} performance synergies often fail to materialize are still poorly understood. Research has explored different paths: First, humans’ ability to exert appropriate reliance depends on the overall decision-making situation, e.g., whether it is possible to ex post verify the correctness of the decisions \citep{Vericourt}. Second, ``imperfections'' on the human side may contribute to this: Humans can struggle to correctly assess their own capabilities in comparison to that of the AI \citep{Fugener2021a}. They may develop implicit biases against the AI that inhibit their willingness to rely on its advice \citep{Turel2023}, or they may start to mirror the AI’s behavior by following its advice even when it is incorrect \citep{Fugener2021}. Third, humans may be misled by signals from the AI that were originally intended to help them better assess its decision quality. Additional explanations may not be correct \citep{Morrison2024}, or they may distort humans’ situational balancing of available information, leading to misconceptions and suboptimal decisions \citep{Bauer2023a}. Overall, further research is required to provide the means for developing a holistic understanding of the synergetic potential between humans and AI---an objective this work pursues by proposing a conceptualization of human-AI complementarity.

\subsection{Concepts Related to Human-AI Complementarity}\label{sec:concepts_related_to_human_AI_complementarity}

Complementarity between humans and AI is discussed as part of several closely related paradigms: \textit{Intelligence augmentation}, \textit{human-machine symbiosis}, and \textit{hybrid intelligence}.

\textit{Intelligence augmentation} is defined as ``enhancing and elevating human’s ability, intelligence, and performance with the help of information technology'' \citep[p. 245]{Zhou2021}. It follows the idea that machines use their abilities to assist humans, not necessarily to achieve CTP, but to improve human objectives. \textit{Human-machine symbiosis} is a paradigm that envisions deepening the collaborative connection between humans and AI. It is based on the notion of a symbiotic relationship between both and considers them as a common system rather than two separate entities with the aim of becoming more effective together than working separately \citep{Licklider1960}. It also makes the assumption that both entities can offer different capabilities that can be leveraged to overcome human restrictions and to reduce the time needed to solve problems \citep{Gerber2020,Jain2021}. \textit{Hybrid intelligence} pursues the idea of combining human and AI team members in the form of a socio-technical ensemble. We refer to the work of \citet[p. 640]{Dellermann2019}, who define hybrid intelligence as ``the ability to achieve complex goals by combining human and artificial intelligence, thereby reaching superior results to those each of them could have accomplished separately, and continuously improve by learning from each other.'' In addition to realizing performance synergies on an individual level, IS research has also considered AI as a performance driver on an organizational level when firms develop an AI capability within the organization \citep{Mikalef2021}.

Nevertheless, existing studies under these labels do not provide theoretical views of human-AI complementarity. Articles by \citet{Donahue2022}, \citet{Steyvers2022}, and \citet{Rastogi2023} are the only works to theorize about human-AI complementarity. \citet{Donahue2022} discuss scenarios in which CTP could occur by considering fairness aspects. \citet{Steyvers2022} derive a framework for combining individual decisions and different types of confidence scores from humans and AI models. Lastly, \citet{Rastogi2023} propose a taxonomy of human and AI strengths together with the notion of across- and within-instance complementarity. All three differ from our approach as they do not distinguish between complementarity potential (\(CP\)) and complementarity effect (\(CE\)) with their respective components. Moreover, they do not empirically analyze human-AI decision-making in behavioral experiments.

\section{Conceptualization of Human-AI Complementarity}\label{sec:concpetualization_human_ai_complementarity_ejis}

In this section, we first introduce the fundamental notion of human-AI complementarity and formalize our decision-making situation as a basis for further analysis. Subsequently, we introduce and formalize the concepts of complementarity potential and effect, and relate them to the underlying sources of complementarity.

\subsection{The Principle of Human-AI Complementarity}\label{sec:principle_human_ai_complementarity_ejis}

\begin{figure}[h]
    \centering
    \includegraphics[width=0.65\textwidth]{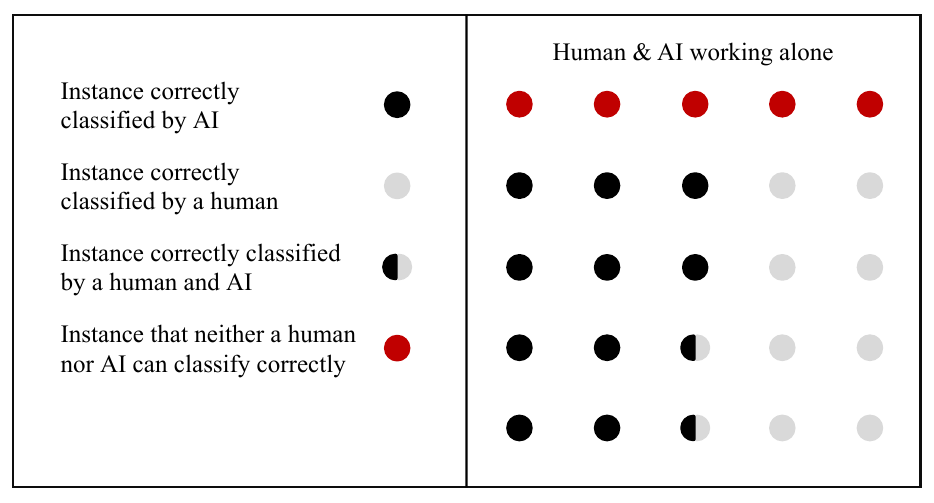}
    \caption{Illustration of the principle of human-AI complementarity based on different (in)correct decisions that human and AI can make.}
    \label{fig:illustration_human_AI_complementarity}
\end{figure}

We first motivate the underlying idea of complementarity which drives effective human-AI decision-making. In this work, we focus on the performance on decision-making tasks that humans and AI can conduct independently---recognizing that AI models have elevated above pure decision support for humans. Real-world examples include diagnosing diseases in medicine \citep{Goldenberg2019}, conducting loan decisions in finance \citep{Turiel2020}, and writing entire programs with AI code generation systems \citep{Ross2023}. However, since neither humans nor AI are perfect, discrepancies in access to information or in capabilities could be leveraged to generate superior outcomes in a human-AI team. \Cref{fig:illustration_human_AI_complementarity} illustrates the situation in a simple example for a set of decisions: The decision-making task comprises 25 instances that each have a set of possible discrete outcomes only one of which is correct. The number of incorrect decisions measures the performance of each individual team member and the human-AI team, respectively. The AI makes 13 incorrect decisions, while the human errs at 15 task instances when conducting the task independently. Five of the instances can neither be solved by the AI nor the human on their own. Consequently, if the human-AI team were just to pick the correct decision of either the AI or the human, the team could correctly solve 20 instances and would only miss the 5 that none of them can solve. In other words, relying on the correct individual decisions of \textit{each} team member improves the result compared to the AI acting alone (as the individually better performing team member): While the AI still shows 13 incorrect decisions when acting independently, teaming up with the human can reduce this number to 5---realizing an improvement potential of 8 decisions. Moreover, it is also conceivable that—while none of the team members is correct individually---the interaction between the team members may allow the generation of correct team decisions even for the remaining 5 red task instances in \Cref{fig:illustration_human_AI_complementarity} representing an additional improvement potential\footnote{We note that this is not possible if team members rely on one of their individual decisions as team decision—as tasks solutions are confined to the solutions provided by each team member alone.}: Team members may recognize through collaboration that they have different information or capabilities that they can jointly apply. Let us, e.g., assume that a task would be to deliver a diagnosis on cancer. Both a human radiologist and an AI analyzing patient data might each individually render a wrong diagnosis, e.g., ``malign cancer'' conjectured by the human and ``no cancer'' proposed by the AI. If, however, the human might learn about the AI’s rationale (e.g., regions in the X-ray scan crucial for the decision) or the AI might learn about ``side'' information that the radiologist has on the patient history, they jointly may arrive at the correct diagnosis ``benign cancer''.

For human-AI teams to achieve CTP, it is essential that they manage to realize these improvement potentials. If their information and capabilities could be adequately combined, the team performance in such situations would be superior to their individual performances \citep{Rastogi2023}.

In the introductory example, performance is captured by the absolute number of errors. Depending on the application context, more intricate measures could also be applied, e.g., precision or recall in classification tasks (e.g., when analyzing radiology images in health care), mean absolute error in prediction tasks (e.g., when making sales forecasts for inventory management), or more complex compound metrics (e.g., when weighting multiple dimensions of interest).

In reality, the performance of humans and AI varies depending on the task and the application domain. Enabled by advances in AI research over the last few years, there has been an increasing number of tasks where the performance of AI has reached or even exceeded that of humans \citep{Afshar2022,Bubeck2023,Silver2018}. However, many applications remain in which human performance remains superior \citep{Brynjolfsson2018,Grace2018,Grace2024}. For the motivating example, we chose a setting in which the AI makes fewer errors—as a reflection of the developments in AI model performance over the last years. However, the conceptualization that we develop in this section is independent of the performance relationship between human and AI.

\subsection{Human-AI Decision-Making Setting}\label{sec:human_ai_decision_making_setting_ejis}

Let us first define the human-AI decision-making setting illustrated in \Cref{fig:illustration_human_AI_complementarity}, which is the foundation of this work: A decision task \(T = \left\{(x^{(i)},\ y^{(i)})\right\}_i^N\) is a set of \(N\) instances \(x^{(i)} \in X\) with corresponding ground truth labels \(y^{(i)}\in Y\) denoting the correct results. In this section, we use the term decision for both classification and prediction tasks. The ground truth, i.e., the correct decision, might not be known at the time of the decision, but can be determined and revealed later. Both a human decision-maker \textit{H} and a machine learning model, which we denote as \textit{AI}, are capable of independently producing a decision for each task instance. For any given instance \(x^{(i)}\), the human and the AI will independently derive decisions \(\hat{y}_H^{(i)}\) and \(\hat{y}_{AI}^{(i)}\). In a scenario where the human and the AI might collaborate in respect of their decision-making, a collaboration mechanism \(I(\hat{y}_H^{(i)}, \hat{y}_{AI}^{(i)})\) integrates their decisions into a final team decision \(\hat{y}_I^{(i)}\). We note that this decision might be different from each individual decision.

Each decision’s quality is measured by its deviation (``loss'') from the ground truth---by a loss function \(l\) bounded in \(R^+\). This function serves as a generic measure of performance that could take different forms with different decision problems, e.g., an error rate used in classification tasks (like the number of wrong or unsolved task instances in \Cref{fig:illustration_human_AI_complementarity}). For any instance \(x^{(i)}\), losses \(l_D^{(i)}\) with \(D \in \left\{H,AI,I\right\}\) are given by \(l_H^{(i)}\) for the human decision, \(l_{AI}^{(i)}\) for the AI decision, and \(l_I^{(i)}\) for the integrated team decision. This results in overall losses \(L_D\) for the entire task by averaging all the available instances:

\begin{equation}
L_D=\frac{1}{N}\sum_{i=1}^{N}{l_D^{(i)}\left({\hat{y}}_D^{(i)},y^{(i)}\right)}\ with\ D\ \in\ \left\{H,AI,I\right\}.
\end{equation}

\subsection{Complementarity Potential}\label{sec:complementarity_potential_ejis}

From a decision-theoretic perspective, the vision of human-AI decision making is to attain a superior team performance compared to the human and the AI conducting the task individually---providing the fundamental reason for forming human-AI teams \citep{Rastogi2023}. In our context, the human-AI team reaches \textit{complementary team performance (CTP)} when the loss of the team is strictly smaller than that of the human and the AI individually \citep{Bansal2021}:

\begin{equation}
CTP = \begin{cases} 
1, & L_I < \min(L_H, L_{AI}), \\ 
0, & \text{otherwise}. 
\end{cases}
\end{equation}

In addition to a binary task outcome, we propose the notion of \textit{complementarity potential (\(CP\))} to measure the discrepancy between the overall loss of the individually better performing team member \(T^\ast \in \{H,AI\}\) with \(L_{T^\ast} = \min\left( L_H, L_{AI} \right)
\) and perfect decisions for all instances of a task, i.e., the selection of the ground truth, with an overall loss of 0:

\begin{equation}
CP=\ L_{T^\ast}=min\left(L_H,L_{AI}\right).
\label{eq:min_L_H_L_AI}
\end{equation}

In our introductory example in \Cref{fig:illustration_human_AI_complementarity}, the overall loss is quantified by the number of wrong decisions. Consequently, the complementarity potential amounts to 13, which is given as the minimum number of individual errors (13 for the AI, 15 for the human).\footnote{For simplification, we report \(L_D\) in the introductory example as the sum instead of the average of all the instances.} It should be noted that only 8 task instances of this potential can be realized by picking the better individual decision, while 5 task instances cannot be solved individually, but only---if at all---in collaboration, resulting in a decision that neither AI nor human could come up with on their own. We reflect this by distinguishing two components: \textit{Inherent} and \textit{collaborative} complementarity potential.

\textit{Inherent} complementarity potential represents improvement potential, i.e., loss reductions, that---from the perspective of the overall more accurate team member \(T^\ast\)---could be contributed by any superior decisions on the instance level by the overall less accurate team member. This means that the team decision is confined to the solutions contributed by one of the team members---in the introductory example (\Cref{fig:illustration_human_AI_complementarity}) represented by the 20 ``gray and black'' task instances solvable by at least human or AI:

\begin{equation}
\hat{y}_I^{(i)} = I(\hat{y}_H^{(i)}, \hat{y}_{AI}^{(i)}) \in \left\{\hat{y}_H^{(i)}, \hat{y}_{AI}^{(i)} \right\}.
\label{eq:integrate_element_yh_yai}
\end{equation}

\textit{Collaborative} complementarity potential, on the other hand, signifies improvement potential, i.e., loss reductions, that go beyond the individual team solutions by generating ``new'' knowledge. It may be noted that this is only possible if there is interaction between the team members enabling them to learn from the result of the partner and realize integrated values \(\hat{y}_I^{(i)} \in Y\) \textit{different} from the individual ones (\(\hat{y}_H^{(i)}\) and \(\hat{y}_{AI}^{(i)}\)). In the introductory example (\Cref{fig:illustration_human_AI_complementarity}), this is captured by the 5 ``red'' task instances not solvable by either team member alone.

\begin{figure}[h]
    \centering
    \includegraphics[width=0.8\textwidth]{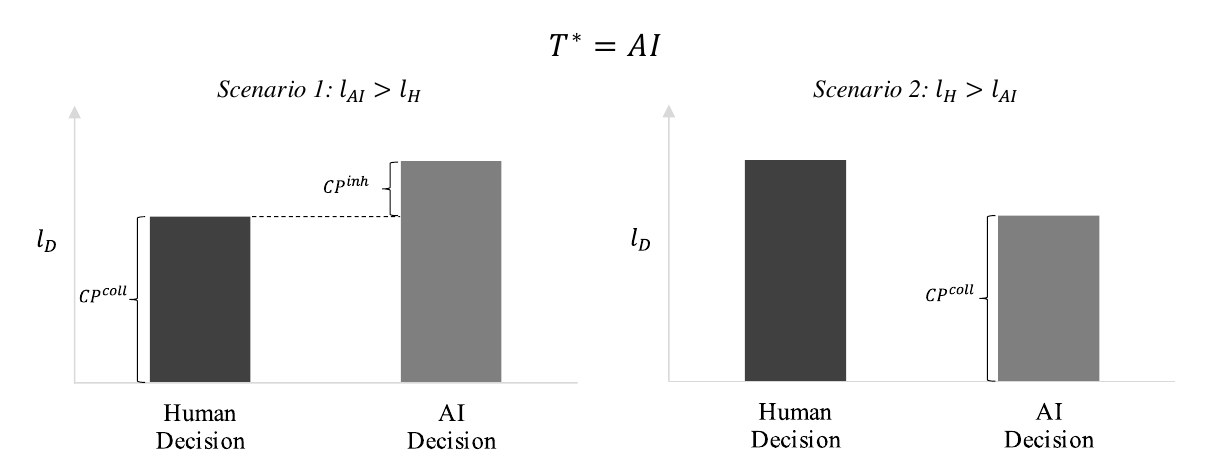}
    \caption{Complementarity potential (\(CP\)) split into inherent (\(CP^{inh}\)) and collaborative (\(CP^{coll}\)) components for a single instance with better human performance (left) and better AI performance (right) in respect of \(T^\ast=AI\). \(l_D\) denotes the instance-specific loss with \(D\in\ \left\{H,AI\right\}\) with a lower loss indicating better performance for the same instance.}
    \label{fig:split_complementarity_potential}
\end{figure}

We illustrate both potentials from a general perspective on a continuous range by looking at the individual losses of human and AI for a single task instance in \Cref{fig:split_complementarity_potential}. Without loss of generality, we assume that the AI is the overall better performing individual team member (\(L_{AI} \leq L_{H}\)). If for this particular instance the overall inferior team member, i.e., the human, could help reduce the loss with his/her decision, we denote \textit{inherent} complementarity potential (scenario 1). The remaining loss is unavoidable if---according to \Cref{eq:integrate_element_yh_yai}---the team decision is restricted to one of the team members’ individual decisions. Thus, the loss of the better performing team member for a task instance constitutes the \textit{collaborative} complementarity potential. This potential could only be exploited if the team members’ collaboration yields new insights for this task instance that were not available for the individual decisions before the collaboration (scenario 1 and 2), resulting in a team decision that incurs a loss lower than that of each team member (human and AI) individually.

Formally, inherent complementarity potential \({CP}^{inh}\) can be calculated by aggregating all potential loss improvements on the instance level where the overall worse performing team member can help improve the team result:

\begin{equation}
{CP}^{inh} = \frac{1}{N} \sum\limits_{i=1}^{N} \begin{cases}
               \max(0, l_{AI}^{(i)} - l_{H}^{(i)}), & L_{AI} \leq L_{H}, \\
               \max(0, l_{H}^{(i)} - l_{AI}^{(i)}), & L_{AI} > L_{H}.
            \end{cases}
\end{equation}

The collaborative complementarity potential \({CP}^{coll}\) can be calculated by aggregating the remaining minimum losses per task instance that the team members incur individually:

\begin{equation}
{CP}^{coll} = \frac{1}{N} \sum_{i=1}^{N} \min(l_H^{(i)}, l_{AI}^{(i)}).
\end{equation}

The inherent and collaborative components are additive and together form the total complementarity potential \(CP\) (see Appendix A for additional details):

\begin{equation}
CP={CP}^{inh}+{CP}^{coll}.
\label{eq:CP_inh_plus_CP_coll}
\end{equation}

In our introductory example, the total complementarity potential of 13 instances can be differentiated into an inherent complementarity potential (\({CP}^{inh}\)) of 8 instances (for which the human team member can contribute the correct solution), and a collaborative complementarity potential (\({CP}^{coll}\)) of 5, with none of the team members arriving at the correct decision individually.

\subsection{Complementarity Effect}\label{sec:complementarity_effect_ejis}

In real-world collaboration scenarios between a human and AI, it is, of course, unlikely that the entire complementarity potential will be exploited. We, therefore, introduce the \textit{complementarity effect} (\(CE\)) as that part of this potential that is actually realized by the integrated team decision. Measuring and dissecting this effect will allow observed human-AI decision-making settings to be analyzed in greater detail, in order to infer conclusions about the collaboration’s effectiveness, and to purposefully develop and compare collaboration designs and mechanisms. Analogous to the complementarity potential in \Cref{eq:min_L_H_L_AI}, the realized complementarity effect accounts for the difference between the average loss of the overall individually better team member and that of the integrated team decision (\(L_I>0\)): 

\begin{equation}
CE\ =\ min(L_H,L_{AI})-L_I.
\label{eq:CE_min_L_H_L_AI_minus_L_I}
\end{equation}

\begin{figure}[h]
    \centering
    \includegraphics[width=0.7\textwidth]{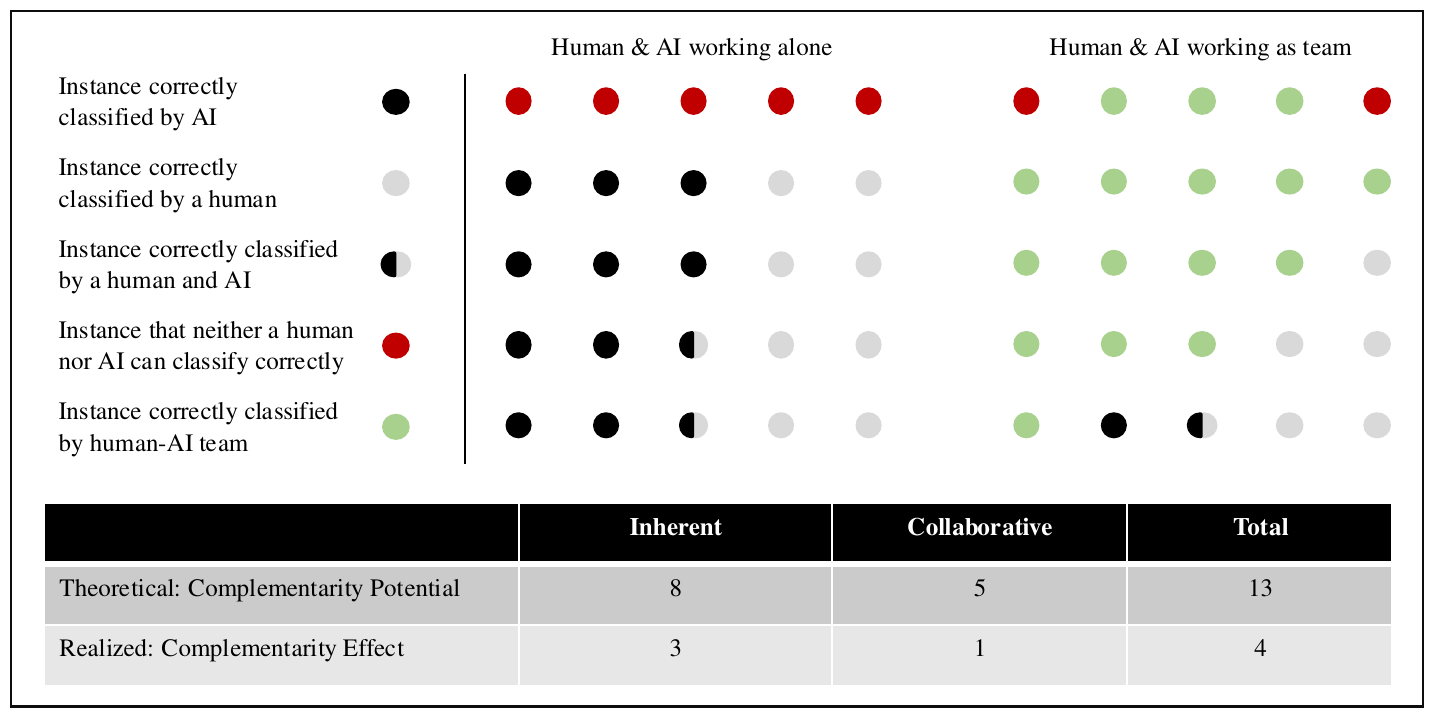}
    \caption{Illustration of (theoretical) complementarity potential and (realized) complementarity effect for a hypothetical situation extending the introductory example in \Cref{fig:illustration_human_AI_complementarity}.}
    \label{fig:illustration_human_AI_complementarity_extended}
\end{figure}

We expand our introductory example (\Cref{fig:illustration_human_AI_complementarity}) in \Cref{fig:illustration_human_AI_complementarity_extended} by incorporating (hypothetical) integrated human-AI decisions for all instances. Let us assume that the human-AI team makes 9 incorrect decisions, compared to 13 errors by the AI and 15 errors by the human when acting independently. Thus, this collaboration generates a complementarity effect of 4---realizing 31\% of the full complementarity potential of 13.

\begin{figure}[h]
    \centering
    \includegraphics[width=0.85\textwidth]{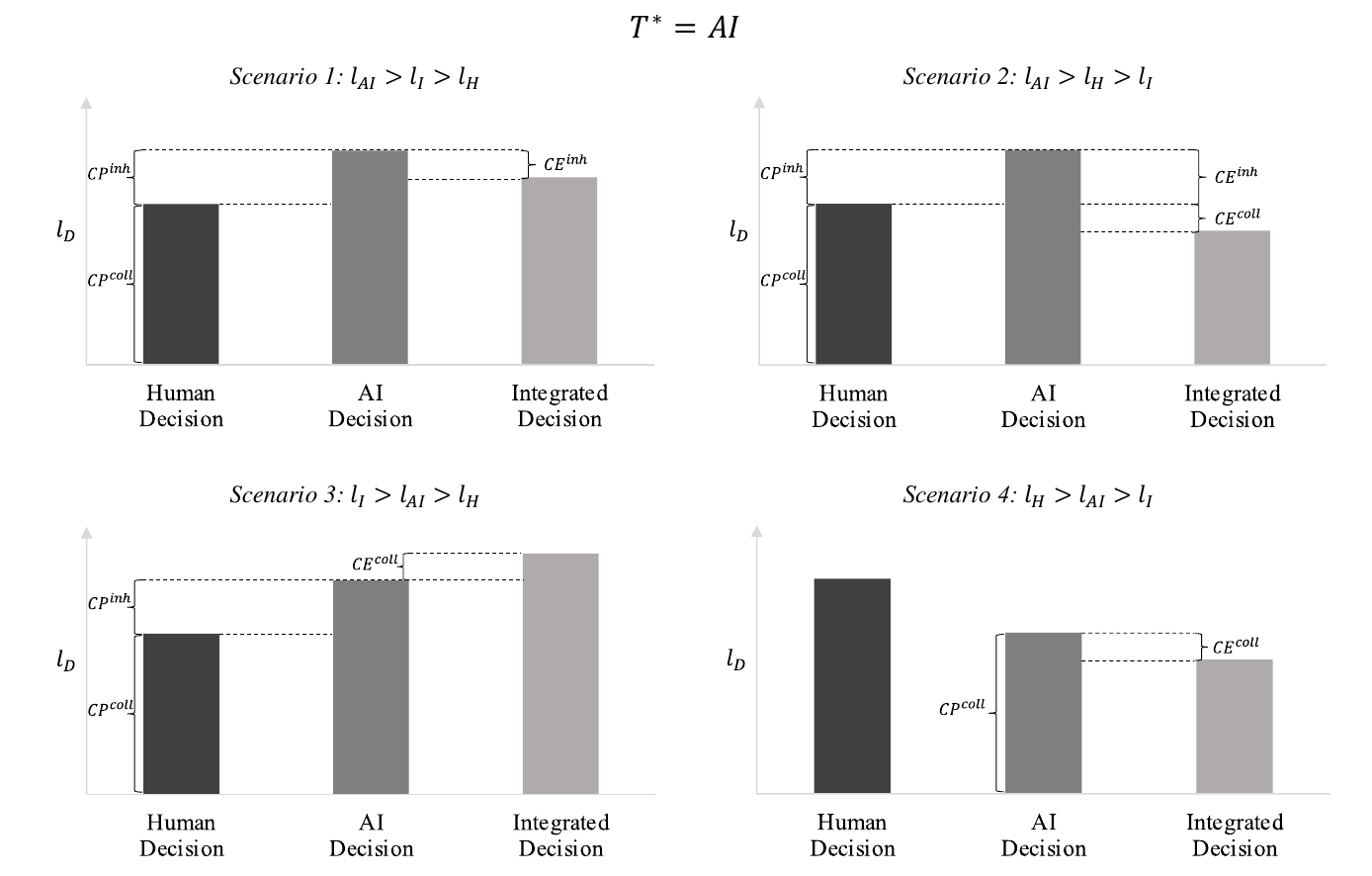}
    \caption{Illustration of (theoretical) complementarity potential (\(CP\)) and (realized) complimentarity effect (\(CE\)) in respect of different loss scenarios for a single instance---assuming, without loss of generality, that the AI performs better overall (\(T^\ast=AI\)). \(l_D\) denotes the instance-specific loss with \(D\in\ \left\{H,AI,I\right\}\)---with a lower loss indicating better performance for the same instance.}
    \label{fig:split_complementarity_effect}
\end{figure}

The complementarity effect measures loss improvements—from the perspective of the overall more accurate team member \(T^\ast\)---that are realized by the integration of individual decisions into a team decision. Analogous to inherent or collaborative complementarity potential, we can split the complementarity effect into the same two categories. \Cref{fig:split_complementarity_effect} illustrates the different scenarios that could materialize for a particular task instance in terms of the losses caused by the solutions of human, AI, and the (integrated) team for the same task instance. Again, we assume, without loss of generality, the AI to be the overall better performing individual team member (\(L_{AI}\le L_H\)). If there is inherent complementarity potential (i.e., the overall inferior human is more knowledgeable for the particular task instance), this potential may be realized either partially (\(l_{AI}>l_I>l_H\), scenario 1), fully (\(l_{AI}>l_H>l_I\), scenario 2), or not at all (\(l_I>l_{AI}\), scenario 3). No inherent complementarity potential exists where the overall better performing team member also dominates in the particular task instance (scenario 4).

In addition, the collaborative complementarity potential (as the smaller loss of the team members in each scenario) could be tapped into. While in scenario 1 only a fraction of the inherent complementarity potential is realized, scenario 2 not only fully exploits the inherent complementarity potential, but also taps into some of the collaborative complementarity potential. Neither scenarios 3 nor scenario 4 realize any inherent complementarity potential, but solely contribute collaborative complementarity effects: In scenario 3 the integrated solution performs even worse than that of the inferior team member—resulting in a \textit{negative} collaborative complementarity effect, i.e., the collaboration actually worsens the outcome. Conversely, in scenario 4, the integrated solution outperforms the better team member, generating a \textit{positive} collaborative complementarity effect.
In general, we can aggregate the complementarity effects across all instances of a task and summarize both cases (AI or human with overall better performance) and the scenarios above (depending on the instance performance of human, AI, and human-AI team):

\begin{align}
{CE}^{inh} &= \frac{1}{N} \sum_{i=1}^{N} \begin{cases}
              l_{AI}^{(i)} - l_I^{(i)}, & L_{AI} \leq L_H \hspace{0.2em}\text{ and } l_{AI}^{(i)} > l_I^{(i)} \geq l_H^{(i)}, \\
              l_{AI}^{(i)} - l_H^{(i)}, & L_{AI} \leq L_H \hspace{0.2em}\text{ and } l_{AI}^{(i)} > l_H^{(i)} > l_I^{(i)}, \\
              l_H^{(i)} \hspace{0.075em}- l_I^{(i)}, & L_H \hspace{0.225em}< L_{AI} \text{ and } l_H^{(i)} > l_I^{(i)} \geq l_{AI}^{(in)}, \\
              l_H^{(i)} \hspace{0.075em}- l_{AI}^{(i)}, & L_H \hspace{0.225em}< L_{AI} \text{ and } l_H^{(i)} > l_{AI}^{(i)} > l_I^{(i)}, \\
              0, & \text{otherwise}.
            \end{cases} \\[10pt]
{CE}^{coll} &= \frac{1}{N} \sum_{i=1}^{N} \begin{cases}
              l_H^{(i)} \hspace{0.075em}- l_I^{(i)}, & l_{AI}^{(i)} \geq l_H^{(i)} > l_I^{(i)}, \\
              l_{AI}^{(i)} - l_I^{(i)}, & l_H^{(i)} \hspace{0.075em}> l_{AI}^{(i)} > l_I^{(i)}, \\
              l_{AI}^{(i)} - l_I^{(i)}, & L_{AI} \leq L_H \hspace{0.225em}\text{ and } l_I^{(i)} > l_{AI}^{(i)}, \\
              l_H^{(i)} \hspace{0.075em}- l_I^{(i)}, & L_H \hspace{0.225em}< L_{AI} \text{ and } l_I^{(i)} > l_H^{(i)}, \\
              0, & \text{otherwise}.
            \end{cases}    
\end{align}

Analogous to \Cref{eq:CP_inh_plus_CP_coll}, the inherent and collaborative components add up to the total complementarity effect (see Appendix A for additional details):

\begin{equation}
{CE=CE}^{inh}+{CE}^{coll}.
\label{eq:CE_inh_plus_CE_coll}
\end{equation}

In our extended introductory example in \Cref{fig:illustration_human_AI_complementarity_extended}, we find that of the 8 task instances offering inherent complementarity potential, 3 could be realized as an inherent complementarity effect (\({CE}^{inh}\)) by taking the human’s individual decision suggestions into account. Of the 5 task instances for which neither the human nor the AI could individually make a correct decision, collaboration enabled the human-AI team to make correct decisions regarding 3 task instances. However, also 2 task instances that the AI alone could have performed correctly are subject to an erroneous decision due to the collaboration. Consequently, the collaborative complementarity effect (\({CE}^{coll}\)) amounts to 1 and the total complementarity effect (\(CE\)) to 4.

\subsection{Sources of Complementarity: The Impact on Complementarity Potential and Effect}\label{sec:sources_of_complementarity_potential_ejis}

In the motivating example, the human and the AI make erroneous decisions for different task instances allowing for possible performance synergies that may be attainable through collaboration. We used this example to introduce the measures of complementarity potential and effect.

Differences in human and AI decision-making can arise from information and capability asymmetries as introduced earlier (\Cref{sec:sources_of_complementarity}). These asymmetries are fundamental for the existence of complementarity. They cause different decisions to be taken by humans and AI on an instance level and, therefore, influence the complementarity potential and the realized complementarity effect including both components. Leveraging information asymmetry means making information available to the team: Advantages can be captured as inherent (when the team decision for a task instance corresponds to the individual decision of the team member with ``better'' information) or as collaborative complementarity potential/effect (when the information of both team members is joined resulting in new insights and a team decision different from the individual ones). Similarly, capability asymmetries may contribute to inherent (when the team decision corresponds to the individual decision of the team member that is more capable for the particular task instance) or collaborative complementarity potential/effect (when capabilities complement each other, e.g., human experience and AI computational power, resulting in a team decision that differs from the individual ones).

It may be noted, though, that realizing complementarity potential (and, thus, achieving CTP) involves managing a number of trade-offs: Information and capabilities are typically not distributed in a way that either the human or the AI dominates across all existing task instances \citep{Geirhos2020,Geirhos2021,Kühl_goutier}. There may be pieces of information that only the human or only the AI has access to: In our earlier radiology example, this may be contextual information about a patient that the physician has and a large variety of cases in the training set that only the AI is able to access. In addition, also information may be traded off against capabilities: Assuming the radiologist has dominating information, it may be outweighed by the capabilities of the AI to automatically digest and evaluate the information available to it in real time.

\begin{figure}[h]
    \centering
    \includegraphics[width=0.78\textwidth]{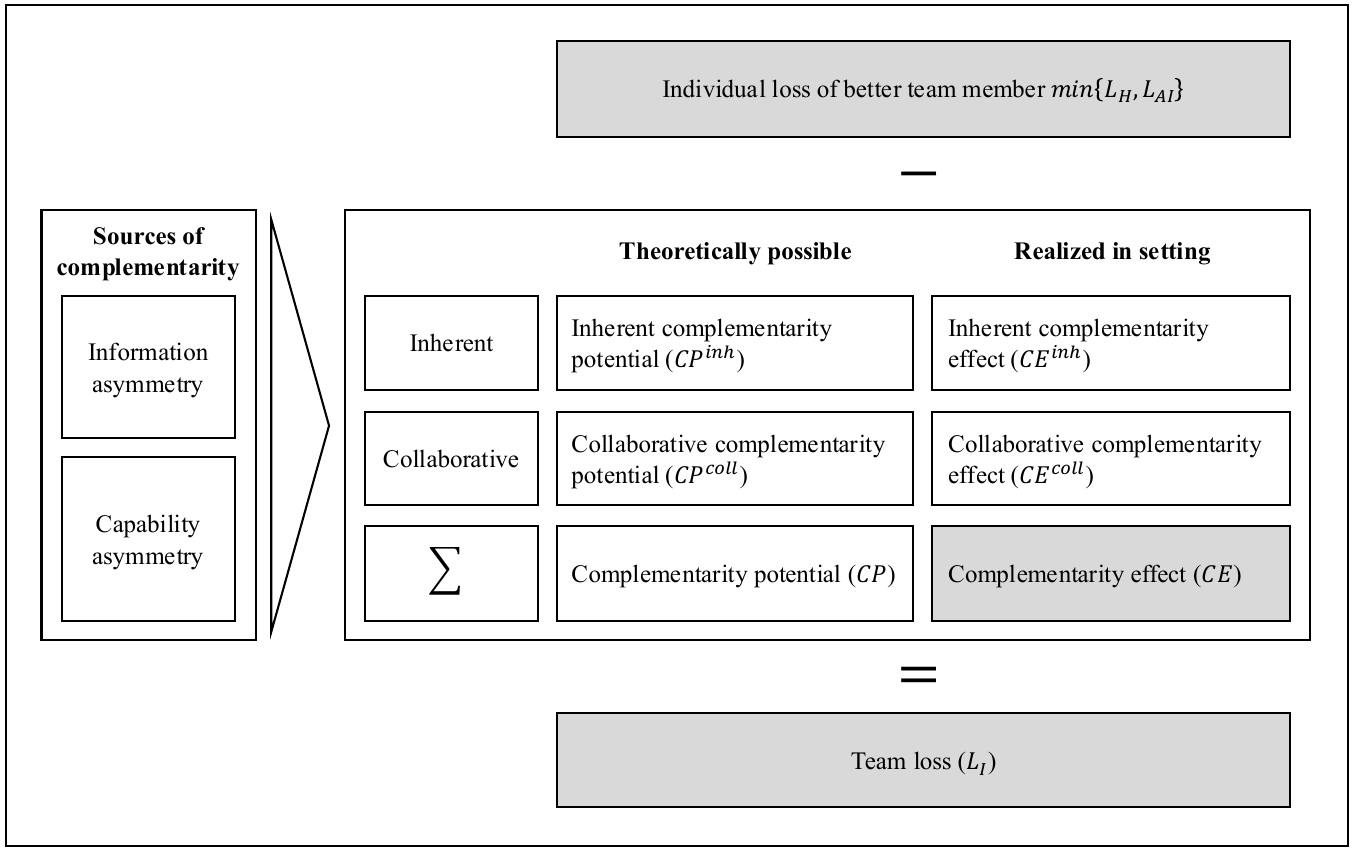}
    \caption{Conceptual framework for human-AI complementarity. It summarizes the complementarity potential and effect, including the inherent and collaborative components. Information and capability asymmetry can affect both the theoretically existing potential and the realized effect.}
    \label{fig:overview_complementarity}
\end{figure}

We summarize the notion of complementarity, its sources, and the measures of complementarity potential and effect in the conceptual framework in \Cref{fig:overview_complementarity}. Each source can affect both the ``inherent'' and ``collaborative'' components of complementarity potential and effect. Our conceptualization presented in this section is intended to provide a deeper understanding of the phenomenon as well as to provide a concrete measurement construct for systematically harnessing complementarity.

To illustrate and evaluate this framework, we now design two behavioral experiments in which humans and AI make decisions on their own and jointly for a set of task instances. We investigate how the presence of information (\Cref{sec:effect_of_information_asymmetry_ejis}) and capability (\Cref{sec:effect_of_capability_asymmetry_ejis}) asymmetry affect the team’s complementarity potential, the realized complementarity effect as well as the joint team performance.

\section{Experimental Studies}\label{sec:experimental_studies_ejis}

To demonstrate the proposed conceptualization’s value and application, and to further investigate human-AI decision-making in the presence of the identified sources of complementarity---information and capability asymmetry---we conducted two behavioral experiments. Specifically, we focused on a team setting in which a human decision-maker has access to AI advice and is subsequently responsible for making the final team decision based on his/her judgment and that of the AI \citep{Green2019}. In this collaboration setup, the team decision either matches the individual human or AI decision---i.e., the human relies fully on his/her judgment or that of the AI—or it can be a function of the individual human and AI decision---i.e., an ``integrated'' decision that potentially differs from both individual ones. In the first experiment, we investigated the effect of information asymmetry on decision-making in the human-AI team in the form of additional contextual information only available to the human. In the second experiment, we investigated the effect of capability asymmetry on joint decision-making with different levels of diversity between the capabilities of humans and AI.

\subsection{Experiment 1: The Effect of Information Asymmetry}\label{sec:effect_of_information_asymmetry_ejis}

In the first experiment, we applied the conceptualization developed in \Cref{sec:concpetualization_human_ai_complementarity_ejis} to study the effect of information asymmetry between humans and AI as a relevant source of complementarity. More precisely, we created an intervention in which humans are given contextual information withheld from the AI to investigate whether and how this affects the final team decision and the realized complementarity effect.

\subsubsection{Task and AI Model}

We drew on a real estate appraisal task provided on the data science website kaggle.com \citep{Kaggle2019}. Since housing is a basic need, and because it is ubiquitous in everyone’s life, all people to some degree have the ability to assess a house’s value on the basis of relevant factors such as size or appearance. The data set encompasses 15,474 houses and contains information about the street, city, number of bedrooms, number of bathrooms, and size (in square feet). In the data set, the house prices denote their listing price. The average house price is \$703,120, ranging between a minimum of \$195,000 and a maximum of \$2,000,000. An image of each house is also provided.

In respect of the house price prediction task, we implemented a random forest regression model as the AI model \citep{Breiman2001}. We drew on the individual trees in the random forest to generate a predictive distribution for each instance and provided the 5\% and 95\% quantiles as indicators of the AI model’s prediction uncertainty. We used 80\% of the data as the training set and 20\% as the test set. We trained the random forest on the following features: the street, city, number of bedrooms, number of bathrooms, and square feet of the house. The house’s image was withheld from the AI model.

In respect of the behavioral experiment, we focused on detached family houses in the test set, with the existing image providing a view of its exterior. From these, we randomly drew a hold-out set of 15 houses to serve as samples for our behavioral experiment. The AI model achieves a performance measured in terms of the mean absolute error (MAE) of \$163,080 regarding the hold-out set, which is representative of its performance on the entire test set. In respect of the condition with unique human contextual information (UHCI), we gave humans an additional image of the house, which is likely to constitute valuable information. Humans are able to leverage their general understanding to form an overall assessment based on the house’s features, the visible surroundings, and its appearance. We conducted an initial pilot study to verify this assumption (Appendix B.1 contains additional details).

\subsubsection{Study Design}

We conducted an online experiment with a between-subject design. We recruited participants from prolific.com. The study included two conditions and randomly assigned each participant to one of these conditions. We did not allow any repeated participation. Each participant passed the following steps (see Figure B.1 in Appendix B.2 for additional details):

\textit{Step 1:} After accepting the task, participants were transferred to our experimental website. They were asked for their consent and to read the instructions. A control question initiated the study.

\textit{Step 2:} Since prior work highlights the importance of task training for the participants, we included a mandatory tutorial \citep{Grootswagers2020}. In order to familiarize the participants with the task and data, both treatments received an identical in-depth introduction to the data set, including summary statistics like the mean and the maximum and minimum house prices as reference points (see Figure B.2 in Appendix B.2). Participants in the treatment without unique human contextual information (i.e., without UHCI) were only given the houses’ tabular data, while the participants in the treatment with unique human contextual information (i.e., with UHCI) were also given images of the house.

\begin{figure}[h]
    \centering
    \includegraphics[width=0.95\textwidth]{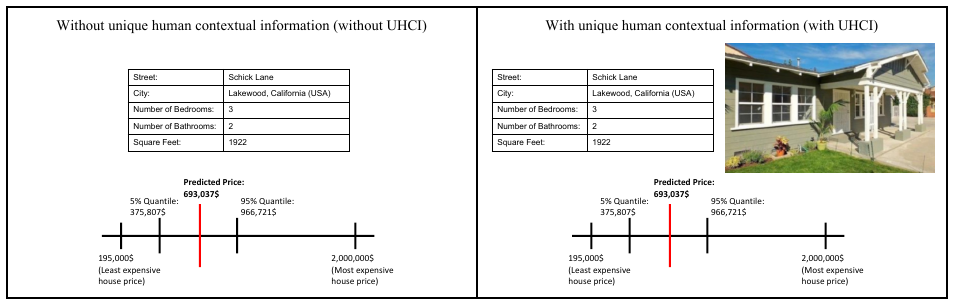}
    \caption{An overview of the interfaces containing the information that the participants were given in the respective behavioral experiment’s treatments.}
    \label{fig:overview_interface_experiment_1}
\end{figure}

\textit{Step 3}: As part of the tutorial, we introduced the participants to the AI. We emphasized that the AI did not have access to the images during the training. We showed the AI prediction in the context of the minimum and the maximum house prices. The participants also received information about the AI’s uncertainty in the form of the 5\% and 95\% quantiles (see \Cref{fig:overview_interface_experiment_1}). We explained the interpretation of the AI’s advice, including all the data points mentioned above (see Figure B.3 in Appendix B.2). The participants were subsequently asked to answer a control question to verify their understanding.

\textit{Step 4:} The participants conducted two training task instances. For each instance, we initially asked the participants to first provide a prediction, before we revealed the AI’s recommendation. They were then asked to adjust the AI’s prediction in the best possible way, constituting the joint human-AI team prediction. After each training example, the participants received feedback in the form of the true house price (see Figure B.4 in Appendix B.2 for an exemplary training task instance). After completing the two training task instances, they were informed about the start of the study.

\textit{Step 5:} Each participant completed 15 house price task instances, presented randomly in the same procedure as described in Step 4 (see Figure B.5 in Appendix B.2 for an exemplary task instance). During the task, the participants were not informed about the actual house price. Subsequently, we asked them to complete a questionnaire regarding qualitative feedback (\textit{Step 6}) and demographic information (\textit{Step 7}).

The overall task lasted approximately 30 minutes. Before recruiting participants, we computed the required sample size in a power analysis using G*Power \citep{Faul2007}. Based on the pilot data, we expected a large effect (\(d =\) 0.8). We referred to an alpha value of 0.05, while taking multiple testing into account in order to achieve a power of 0.8. This resulted in a total sample size of 86. Anticipating that some participants will fail the attention checks, we recruited a total of 120 participants (60 per condition). They received a base payment of £5 and are additionally incentivized following the approach of \citet{Kvaløy2015}, who show the benefits of combining non-monetary motivators, such as recognition, attention, and verbal feedback, with performance-based pay. We achieved this by adding motivational statements and by giving the top 10\% participants an additional pound. Note that the two training tasks are not included in the final evaluation. To ensure the quality of the collected data, we removed those participants whose entered prices exceed the communicated maximum house price of \$2,000,000 in the data set. We also identified outliers for removal by using the median absolute deviation \citep{Leys2013,Rousseeuw1993}. After applying these criteria, we continued with the data of 101 participants across both conditions—53 in the treatment without UHCI and 48 in that with UHCI (see Table B.1 in Appendix B.3 for additional details about the participants).

\subsubsection{Evaluation Measures}

For each participant, we measured the loss of the human (\(l_H\)), the AI (\(l_{AI}\)), and the team decision (\(l_I\)) as the absolute error, and calculated the average over all task instances to receive the human (\(L_H\)), the AI (\(L_{AI}\)), and the team performance (\(L_I\)) corresponding to the mean absolute error (MAE). Furthermore, we calculated the complementarity potential’s and complementarity effect’s respective components as defined in \Cref{sec:concpetualization_human_ai_complementarity_ejis}. Finally, for each measure, we calculated the average over all the participants.

\subsubsection{Results}\label{sec:results_experiment_1}

In this section, we analyze the impact of unique human contextual information on the team performance, the complementarity potential, and the complementarity effect. We evaluate the results’ significance by using the Student’s T-test and the Mann-Whitney U-test, depending on whether the prerequisites have been fulfilled. We apply the Bonferroni correction and adjust the p-values accordingly. First, we focus on the impact of contextual information on performance, followed by an in-depth analysis of its impact on complementarity potential and effect.

\begin{figure}[h]
    \centering
    \includegraphics[width=0.65\textwidth]{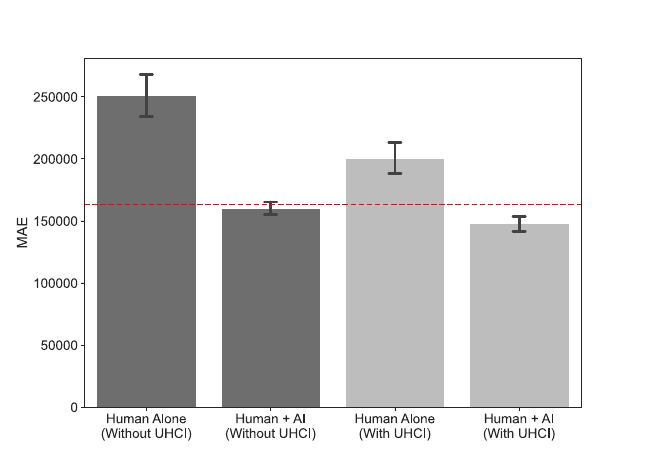}
    \caption{Performance results as the MAE across the conditions (UHCI \(=\) unique human contextual information), including 95\% confidence intervals. The red horizontal line denotes the AI performance.}
    \label{fig:MAE_across_conditions}
\end{figure}

\Cref{fig:MAE_across_conditions} displays the isolated human and joint human-AI performance for both conditions. It also includes the performance of the AI alone. We first evaluate the impact of unique human contextual information without any AI assistance. Participants in the treatment without UHCI achieve an MAE of \$251,282, while those in the treatment with UHCI yield an MAE of \$200,510---an improvement of \$50,772 (20.21\%), which is significant (\(d =\) 0.92, \(p <\) 0.001, two-sample, two-tailed T-test). This result confirms the general usefulness of the provided house images from the human perspective.

Next, we evaluate the impact of unique human contextual information when the human is teamed with the AI. The team performance in the treatment without UHCI results in an MAE of \$160,095 versus an MAE of \$148,009 in the treatment with UHCI—an improvement of \$12,086 (7.55\%), which is significant (\(d =\) 0.59, \(p <\) 0.05, two-sample, two-tailed T-test). In both treatments, the human-AI team outperforms the AI (MAE: \$163,080). Whereas the difference between the performance of the human-AI team and the performance of the AI alone is significant in the treatment with UHCI (\(d =\) 0.68, \(p <\) 0.001, one-sample, two-tailed T-test), the difference in the treatment without UHCI does not constitute a significant improvement (\(d =\) 0.16, \(p =\) 1.0, one-sample, two-tailed T-test).

\paragraph{Complementarity potential.} First, we analyze the inherent complementarity potential (\({CP}^{inh}\)). We observe a significant increase due to the unique human contextual information. In the condition without UHCI, the \({CP}^{inh}\) is \$42,995, and increases to \$61,970 in the condition with UHCI (\(d =\) 1.05, \(p <\) 0.001, two-tailed Mann-Whitney U test). This finding can be interpreted that the images indeed contain useful contextual information for humans, which the AI cannot access, resulting in fewer shared errors that humans and AI make individually.

Next, we calculate the collaborative complementarity potential (\({CP}^{coll}\)). Whereas in the condition without UHCI, the \({CP}^{coll}\) results in \$120,085, in the condition with UHCI it amounts to \$101,110. The difference is statistically significant (\(d =\) 1.05, \(p <\) 0.001, two-tailed Mann-Whitney U test). Since the participants in both conditions work with the same AI model, which has an overall better individual performance, the \(CP\) (i.e., the sum of the inherent and collaborative component) is \$163,080 in both conditions. As \({CP}^{inh}\) increases due to UHCI, \({CP}^{coll}\) decreases because the \(CP\) remains constant.

\paragraph{Complementarity effect.} Then, we focus on the realized complementarity potential, i.e., the complementarity effect (\(CE\)). We find a significant difference between the inherent complementarity effect (\({CE}^{inh}\)) in both conditions (without UHCI: \$14,468; with UHCI: \$27,860; \(d =\) 0.87, \(p <\) 0.001, two-tailed Mann-Whitney U test), which highlights contextual information’s potential. This absolute increase might be due to an increase in inherent complementarity potential and/or an improvement in the integration of both team members’ predictions through the human. In order to investigate this further, we also calculate the inherent complementarity effect’s (\(\frac{{CE}^{inh}}{{CP}^{inh}}\)) relative amount. This analysis reveals that unique human contextual information not only enhances the theoretically available inherent complementarity potential, but that the participants could also use significantly more of it (without UHCI: 34\%; with UHCI: 45\%; \(d =\) 0.82, \(p <\) 0.001, two-tailed Mann-Whitney U test). This is an interesting result, as having more information available than the AI might also have detrimental psychological effects. \citet{Jussupow2020}, for example, find that perceived AI abilities and expertise influence aversion towards AI. This could lead to humans taking less account of the AI’s suggestions when making their final decision \citep{Longoni2019,Mahmud2022,Sieck2005}, which could result in CTP not being achieved. However, our results show that unique human contextual information does not only increase the potential, but also the integration’s overall effectiveness.

\begin{figure}[t]
    \centering
    \includegraphics[width=0.75\textwidth]{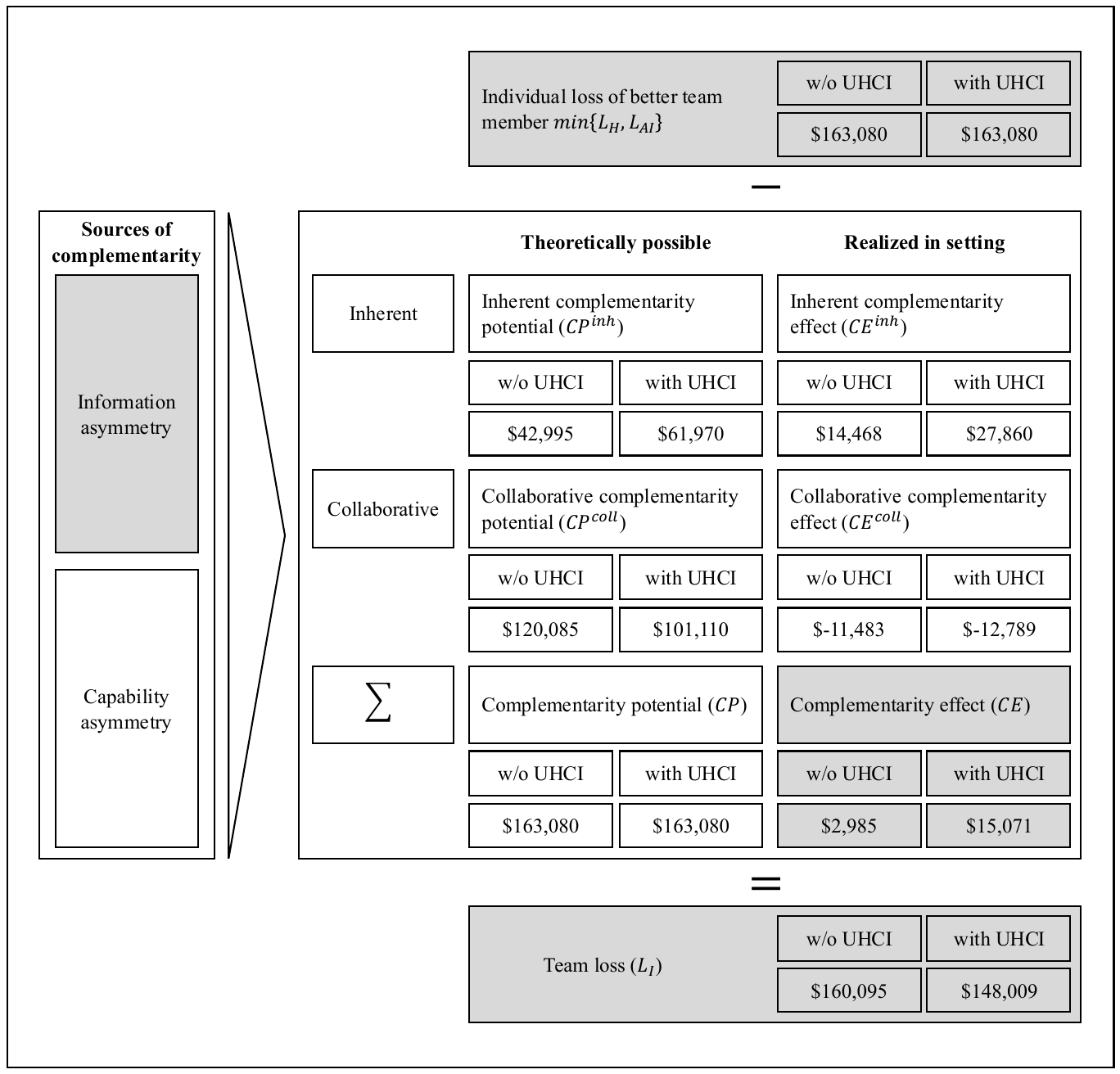}
    \caption{Result summary of the real estate appraisal experiment.}
    \label{fig:summary_real_estate_appraisal}
\end{figure}

Next, we analyze unique human contextual information’s impact on the collaborative complementarity effect (\({CE}^{coll}\)). We do not find a significant difference between the two treatments (\({CE}^{coll}\): without UHCI: \$-11,483; with UHCI: \$-12,789; \(d =\) 0.08, \(p =\) 1.0, two-tailed Mann-Whitney U test). In our experiment, it seems intuitive for the \({CE}^{coll}\) to yield a negative value. Given that the AI in our setup outperforms its human team member, a positive \({CE}^{coll}\) could only occur on task instances where the team loss is even lower than that of the AI and the human (see \Cref{fig:split_complementarity_effect}). Conversely, each instance where the human underperforms in comparison to the AI and fails to improve the AI’s decision contributes negatively to the \({CE}^{coll}\). Since humans tend to choose decisions between two boundaries (in our case their own and the AI’s decision), our setup naturally fosters a negative collaborative complementarity effect. Nevertheless, investigating the individual performance of humans and AI as well as the team performance for each house separately reveals that the human-AI team can derive team decisions for task instances 1, 4, and 13 that are, on average over all participants, more accurate compared to the respective individual decisions of human and AI (see Table B.2 in Appendix B.4). This demonstrates the occurrence of a positive collaborative complementarity effect for these three task instances. See Appendix B.4 for further results regarding a performance analysis of each house.

Overall, the most important finding is that humans can realize a disproportionally large amount of the inherent complementarity potential through unique contextual information, which finally results in CTP. This constitutes a new empirical insight that could only be measured due to the granular formalization. Summing \({CE}^{inh}\) and \({CE}^{coll}\) results in the total complementarity effect (\(CE\)), which equals the performance difference between the best individual team member and the joint human-AI team performance (without UHCI: \$2,985; with UHCI: \$15,071). \Cref{fig:summary_real_estate_appraisal} summarizes the results of our experiment.

\subsection{Experiment 2: The Effect of Capability Asymmetry}\label{sec:effect_of_capability_asymmetry_ejis}

In the second behavioral experiment, we applied the conceptualization to study the effect of capability asymmetry between humans and AI as another relevant source of complementarity. In detail, starting with a ``baseline'' AI, we created an intervention in which we increase the asymmetry between the AI’s and the humans’ capabilities while keeping its overall performance constant. This means the second AI tends to make correct decision suggestions for task instances that tend to be more difficult for humans (``complementary'' AI).

\subsubsection{Task and AI Model}

To investigate the effect of capability asymmetry on the human-AI team performance in decision-making, we chose the image recognition context. Research has demonstrated that humans and AI tend to make different errors on image classification tasks \citep{Fugener2021,Steyvers2022}. Specifically, an AI model based on deep convolutional neural networks tends to infer classification decisions differently than humans do \citep{Geirhos2020,Geirhos2021}. We could therefore expect the AI model to classify certain images more accurately than humans and vice versa, thereby creating inherent complementarity potential. However, it remains unclear whether this naturally existing potential could be sufficiently realized when humans incorporate the AI decision into a final team decision, and whether its increase in the intervention affects the realization.

In order to undertake the experiment, we drew on the image data set that \citet{Steyvers2022} provided. The data set comprises 1,200 images distributed evenly across 16 classes (e.g., airplane, dog, or car). It was curated on the basis of the ImageNet Large Scale Visual Recognition Challenge (ILSVRC) 2012 database \citep{Russakovsky2015}. To increase the task difficulty for humans and the AI, the authors applied phase noise distortion at each spatial frequency, which was uniformly distributed in the interval \([-\omega,\omega]\) with \(\omega = \) 110. Despite the heightened difficulty level, both humans and AI can attain comparable performance on the task. In addition to ground truth labels, the data set also contains multiple human predictions for each image provided by crowd workers, allowing us to infer a proxy for human classification difficulty. Images with a high disagreement in respect of multiple human predictions indicate a higher level of difficulty for humans.

We implemented the AI model as a convolutional neural network, more precisely, as a DenseNet161 \citep{huang2017}, pre-trained on ImageNet. We partitioned the data set into a training (60\%), validation (20\%), and test set (20\%). In the baseline condition, we fine-tuned the AI model on the distorted images over 100 epochs, applying early stopping on the validation loss. We used SGD as an optimizer with a learning rate of \(1\cdot{10}^{-4}\), a weight decay of \(5\cdot{10}^{-4}\), a cosine annealing learning rate scheduler, and a batch size of 32. The AI model achieves a classification error of 26.66\% on the test set.

In the intervention, we created an alternative AI model that makes erroneous decisions for different instances. We fine-tuned the DenseNet161 model exactly as in the baseline condition, but, for each image in the training set, also incorporated a human prediction as an additional label in the training process in order to incentivize the AI model to learn to correctly classify the images that tend to be more difficult for humans \citep{Hemmer2022,Madras2018,Wilder2020}. See Appendix C.1 for additional implementation details of this approach. Even though the AI model has a slightly higher classification error of 33.75\%, this approach results in higher capability asymmetry, i.e., non-overlapping capabilities between the humans and the AI model. We selected 15 images from the test set for the experiment, such that both AI models exhibit the same performance on the test set (26.66\%), while considering non-overlapping errors between both AI models.

\subsubsection{Study Design}

We conducted an online experiment, employing a between-subject design with two conditions. Participants were recruited from prolific.com and randomly assigned to one of the conditions; repeated participation was not allowed. We employed a similar experimental set up as in the first experiment (see Figure C.1 in Appendix C.2 for additional details):

\textit{Step 1:} The participants were transferred to the experimental website after accepting the task, after which they had to submit a consent form and answer an initial control question. Thereafter they had to pass an attention check.

\textit{Step 2:} They were given a task tutorial which presents an exemplary image along with the 16 classes arranged in a four-by-four matrix, including the 16 class icons with their name displayed underneath (see Figure C.2 in Appendix C.2).

\textit{Step 3:} The participants were given an introduction to the AI and its decision regarding an exemplary image followed by a control question to verify their understanding (see Figure C.3 in Appendix C.2). \Cref{fig:overview_interface_experiment_2} displays the interface that the participants see throughout the experiment when collaborating with the AI. In this experiment, the participants were only shown the AI’s decision, since confidence information generated by two separately trained AI models would have introduced a confounder. 

\begin{figure}[h]
    \centering
    \includegraphics[width=0.55\textwidth]{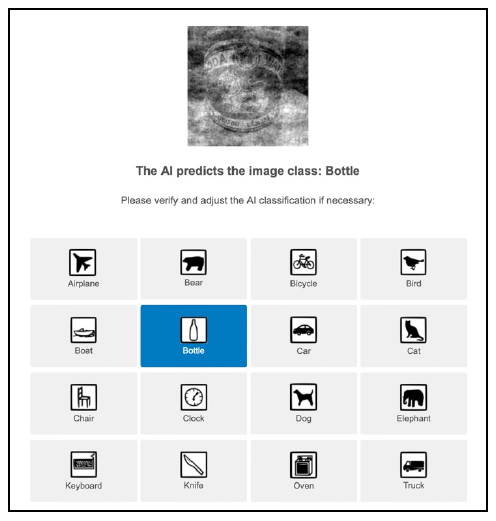}
    \caption{An overview of the interfaces that the participants are shown in both treatments of the behavioral experiment.}
    \label{fig:overview_interface_experiment_2}
\end{figure}

\textit{Step 4:} Next, participants started a practice round comprising three images that had to be classified without AI support in order to familiarize themselves with the classification task (see Figure C.4 in Appendix C.2 for an exemplary training task instance).

\textit{Step 5:} After being informed about the start of the main task, each participant classified 15 images in a randomized order (see Figure C.5 in Appendix C.2 for an exemplary task instance). The participants provided their own classification for each image, then received the AI recommendation and were asked to verify and adjust, if required, the AI classification in the best way possible. During the task, they were not informed about the true class of any image. After classifying all the images, participants were asked to answer a questionnaire regarding qualitative feedback (\textit{Step 6}) and demographic information (\textit{Step 7}).

The overall task lasted approximately 20 minutes. Before recruiting participants, we computed the required sample size in a power analysis using G*Power \citep{Faul2007}. We tested for a medium to large effect (\(d =\) 0.65) and considered an alpha value of 0.05, while taking multiple testing into account in order to achieve a power of 0.8. This resulted in a total sample size of 128. In order to buffer for participants potentially failing attention checks, we recruited a total of 170 participants. They received a base payment of £8 and were additionally incentivized following the approach pursued in the first behavioral experiment \citep{Kvaløy2015}. We excluded participants who did not pass the integrated attention checks, resulting in 144 participants---76 in the base condition (baseline AI) and 68 in the intervention (complementary AI). We provide further details in Table C.1 in Appendix C.3.

\subsubsection{Evaluation Measures}

For each participant, the loss of the human (\(l_H\)), the AI (\(l_{AI}\)), and the team decision (\(l_I\)) was measured as the classification error and averaged over all the task instances, providing the human (\(L_H\)), the AI (\(L_{AI}\)), and the team performance (\(L_I\)). We also calculated the complementarity potential and effect, including their components (see \Cref{sec:concpetualization_human_ai_complementarity_ejis}). Finally, for each measure, we calculated the average over all the participants.

\subsubsection{Results}

We analyze the effect of capability asymmetry on team performance, complementarity potential, and complementarity effect while assessing the statistical significance by using the same procedure as in the first experiment.

\begin{figure}[h]
    \centering
    \includegraphics[width=0.65\textwidth]{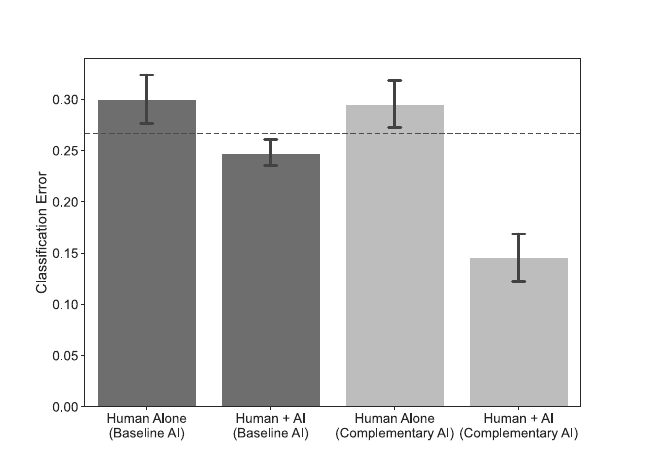}
    \caption{Performance results as classification error across conditions, including 95\% confidence intervals. The red horizontal line denotes the AI performance.}
    \label{fig:classification_error_across_conditions}
\end{figure}

\Cref{fig:classification_error_across_conditions} shows the classification error for humans performing the task alone and together with the AI in both conditions. In addition, it also includes the classification error of both AI models, which are identical in this task. Humans conducting the task alone exhibit a classification error of approximately 0.30, which is nearly identical across the conditions (Baseline AI: 0.2999; Complementary AI: 0.2951; \(d =\) 0.05, \(p =\) 1.0, two-sample, two-tailed T-test). When humans are teamed with the AI, the joint performance increases in both conditions. Whereas the human-AI team yields a classification error of 0.2473 in the condition with the baseline AI, this error decreases even further to 0.1461 in the team with the complementary AI. This corresponds to an improvement of 41\%, which is significant (\(d =\) 1.29, \(p <\) 0.001, two-sample, two-tailed T-test). Both classification errors are significantly lower than that of the AI conducting the task alone in both conditions (Baseline AI: 0.2666, \(d =\) 0.33, \(p <\) 0.05, one-sample, two-tailed T-test; Complementary AI: 0.2666, \(d =\) 1.25, \(p <\) 0.001, one-sample, two-tailed T-test).

\paragraph{Complementarity potential.} We observe a significant increase in the inherent complementarity potential (\({CP}^{inh}\)) in the condition with the complementary AI (Baseline AI: 0.0640, Complementary AI: 0.2480; \(d =\) 3.81, \(p <\) 0.001, two-tailed Mann-Whitney U test). This reflects a higher level of capability asymmetry between humans and the AI. Compared to the baseline AI condition, the AI makes more erroneous decisions for instances that humans can process correctly, whereas humans err for instances that AI can classify correctly. Conversely, we observe a significant decrease in the collaborative complementarity potential (\({CP}^{coll}\)) in the condition with the complementary AI (Baseline AI: 0.2026, Complementary AI: 0.0186, \(d =\) 3.81, \(p <\) 0.001, two-tailed Mann-Whitney U test) as the \(CP\) remains constant due to the AI being individually more accurate than the humans. Whereas the inherent complementarity potential constitutes 24\% of the overall complementarity potential in the baseline condition, this share rises to 93\% in the complementary AI condition. This means that, for the majority of the task instances, one team member is theoretically capable of making a correct decision.

\paragraph{Complementarity effect.} In the baseline condition, 58\% of the inherent complementarity potential (\(\frac{{CE}^{inh}}{{CP}^{inh}}\)) could be realized by the humans integrating their own decision and that of the AI into a final team decision, resulting in a \({CE}^{inh}\) of 0.0368. In the condition with the complementary AI, it is possible to realize 89\% of the inherent complementarity potential, resulting in a \({CE}^{inh}\) of 0.2196. This shows a significant performance improvement (\(d =\) 3.43, \(p <\) 0.001, two-tailed Mann-Whitney U test), which is attributable to a significantly larger fraction of \(\frac{{CE}^{inh}}{{CP}^{inh}}\) that could be realized (\(d =\) 1.02, \(p <\) 0.001, two-tailed Mann-Whitney U test). It indicates that humans tended to rely on the AI decisions when they were correct, but on their decision when it was incorrect. Moreover, in both conditions the collaborative complementarity effect (\({CE}^{coll}\)) is negative. Whereas the value is only slightly negative in the baseline condition (Baseline AI: -0.0175), it decreases to -0.0990 in the condition with the complementary AI. The difference between the two conditions is significant (\(d =\) 1.32, \(p <\) 0.001, two-tailed Mann-Whitney U test).

\begin{figure}[t]
    \centering
    \includegraphics[width=0.75\textwidth]{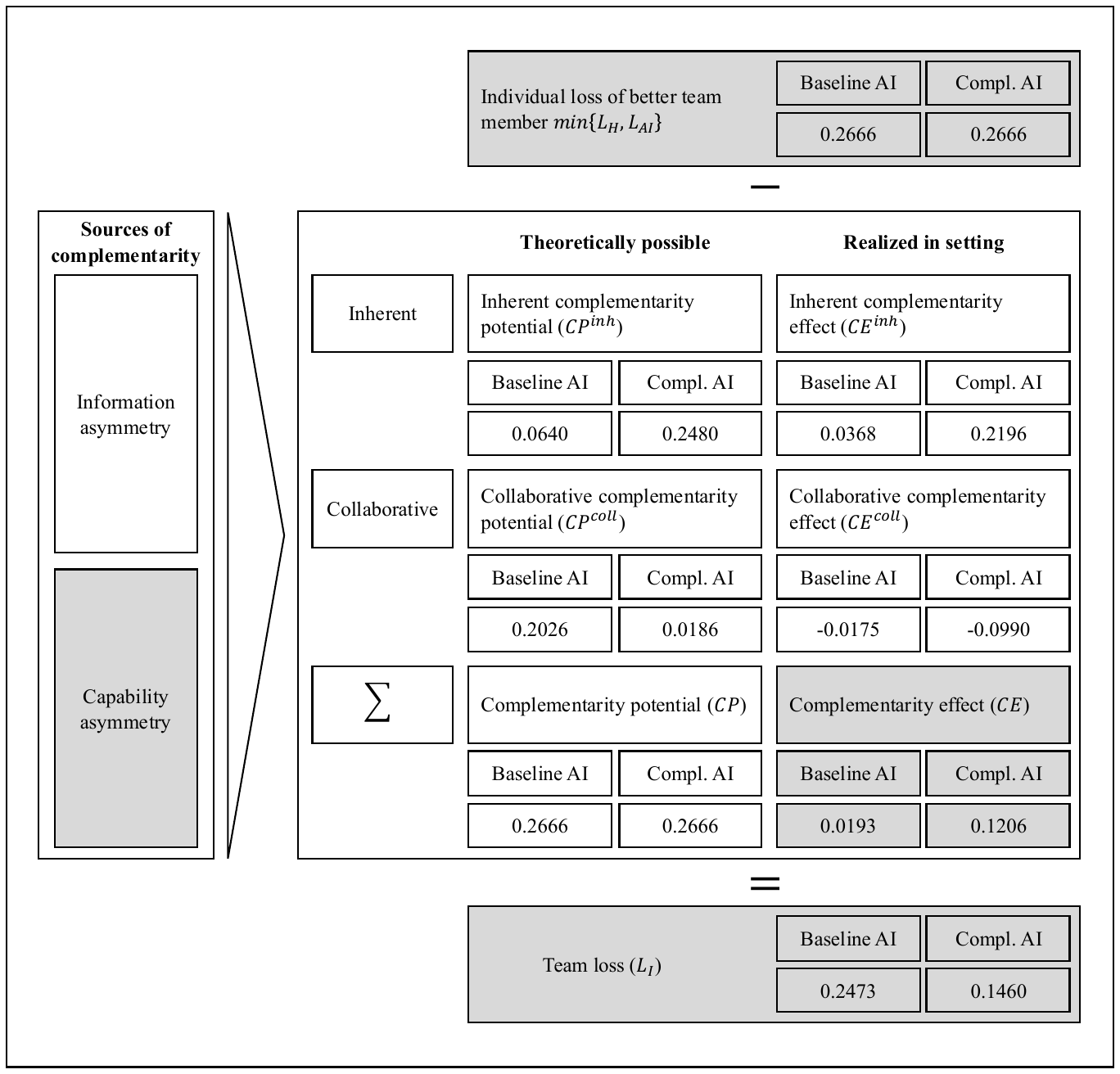}
    \caption{Result summary of the image classification experiment.}
    \label{fig:summary_image_classification}
\end{figure}

The observed changes in both components reveal interesting insights. The increase in the inherent complementarity effect reveals that humans continue to regard the AI as a team member they can rely on, especially with regard to the instances that they find difficult, even though they have observed the AI making several incorrect decisions in respect of the instances that they find easier. Conversely, the decrease in the collaborative complementarity effect reveals that there are also instances where humans, incorrectly, do not rely on the AI. This is an interesting finding, since it could have been conceivable for humans to avoid AI suggestions to a greater extent after witnessing incorrect AI decisions \citep{Dietvorst2015}. Consequently, in the team setting in which the human integrates his/her own decision and that of the AI into a final team decision, there might be a trade-off between turning a higher level of capability asymmetry into performance synergies and not relying on the AI’s suggestions due to witnessing erroneous decisions in ``easier'' instances. We refer to Appendix C.4 for additional analyses, including an analysis of each image. In this context, similar to the first behavioral experiment, we find a positive collaborative complementarity effect which occurs for individual participants for three task instances (see Table C.4 in Appendix C.4).

In summary, \({CE}^{inh}\) and \({CE}^{coll}\) result in the total complementarity effect (\(CE\))---equivalent to the performance difference between the best individual team member and the team performance (Baseline AI: 0.0193, Complementary AI: 0.1206). \Cref{fig:summary_image_classification} summarizes the analysis.

\section{Discussion}\label{sec:discussion_ejis}

Our research emphasizes the importance of complementarity and attaining CTP in human-AI decision-making as an important area of human-AI collaboration \citep{Lai2021}. In the past, when early AI-based systems still had limited capabilities, CTP was rarely a focus, as these systems could only provide humans with (partial) decision support within a small range of low-stakes decision-making tasks, e.g., calculating decision-making inputs \citep{Power2008}. Today, however, AI can perform a growing number of tasks independently with human-level performance and is increasingly utilized in high-stakes decision-making domains such as medicine \citep{McKinney2020}, law \citep{Hillman2019}, and finance \citep{Day2018}. It can also provide general-purpose support, driven by recent advances in large language models that enable applications like ChatGPT \citep{Bubeck2023}. While this opens up the potential for automation, it also enables a new form of collaboration in which AI and humans are equitable team members.

\subsection{Contributions}

Current research lacks a concise concept of human-AI complementarity, as the majority of studies focusing on human-AI collaboration in decision-making do not achieve CTP \citep{Bansal2021,Hemmer2021}. Moreover, merely observing whether CTP has been achieved or not does not allow in-depth conclusions to be drawn about the synergetic potential of humans and AI through collaboration. In this work, we contribute to a comprehensive understanding of human-AI teams’ inner workings regarding decision-making and provide support on how to achieve CTP more consistently.

More specifically, we develop a conceptualization of human-AI complementarity that introduces and formalizes the notion of complementarity potential and complementarity effect and outlines relevant sources of complementarity. This conceptualization allows for analyzing human-AI teams’ potential for decision-making synergies by making this potential measurable. By differentiating between the inherent and collaborative components of complementarity potential and its realized effect, it is also possible to gain important insights into the collaboration’s functioning in terms of the optimal joint team decision. These insights, along with the sources of complementarity, can inform human-AI teams’ future designs and collaboration mechanisms, depending on the use case and application domain. Furthermore, we not only demonstrate the conceptualization’s utility in two behavioral experiments with humans in the role of integrating the final team decision, but we also empirically show that both sources of complementarity---information and capability asymmetry---can actually lead to CTP. In this context, our conceptualization allows us to reveal interesting empirical insights. The first experiment highlights that providing humans with unique contextual information not only affects the inherent complementarity potential, but can also disproportionally increase the realized amount, i.e., the inherent complementarity effect. This means that the effectiveness of the integration that humans conducted improved, which is an interesting finding. Intuitively, the perception of having more information than the other team member could alternatively result in reduced utilization of the AI suggestions in the team prediction—resulting in a sub-optimal team performance \citep{Jussupow2020,Mahmud2022,Sieck2005}. The second experiment reveals that if the human-AI team’s capability asymmetry is greater, this can contribute to the realization of a larger share of inherent complementarity potential. This is also an insightful observation, since humans observing erroneous AI decisions for instances that they could solve relatively easily themselves might lead them to refute AI support even in situations when it is actually helpful \citep{Dietvorst2015}. This finding also contributes to our understanding of team diversity’s compositional impact on performance \citep{Horwitz2005}. Whereas expertise diversity in human teams could be a performance driver, e.g., because it fosters a broader range of cognitive skills \citep{Cohen1997}, it could also turn into a performance inhibitor, e.g., due to the potential difficulties of achieving a mutual understanding \citep{Dougherty1992}. As in human teams, there might also be a trade-off, since humans could start avoiding AI suggestions if they observe them to be incorrect too often \citep{Dietvorst2015}. Our conceptualization enables an analysis of this trade-off---finding that, in the specific experiment, the capability asymmetry between humans and AI affects team performance positively. Finally, the results of both experiments provide empirical evidence for the validity of our conceptualization. They empirically support the existence of both the inherent and the collaborative components. In particular, we find that not only are humans able to realize the inherent complementarity potential in the team, but also that the interaction between humans and AI can lead to more accurate team decisions for several task instances compared to their individual decisions, thus proving the existence of the collaborative component.

\subsection{Theoretical Implications}

From a theoretical perspective, the proposed conceptualization provides a foundation for future human-AI collaboration research. We offer the research community concrete measures that allow the investigation of human-AI teams’ inner workings at a deeper level than aggregated performance comparisons in behavioral experiments allow. In this context, these measures could also help researchers formulate and test hypotheses about the effect of different socio-technical factors on the realization of decision-making synergies in human-AI teams \citep{Jain2021}.

Our research’s most important implication is the need to design for CTP, which is influenced by the source of complementarity and the collaboration mechanism to derive a team decision. Both should be taken into account purposefully. Increasing unique knowledge in the team positively affects the inherent complementarity potential. From an AI perspective, this could be achieved by deliberately creating ``complementary'' AI models designed for teamwork that perform well in areas of the feature space where humans experience difficulties \citep{Hemmer2022,Mozannar2020,Wilder2020}. From a human perspective, humans could be trained to focus on their unique capabilities and to build awareness to use unique contextual information. Over time, the capabilities of humans and AI may also change. While humans become more experienced, AI models may get updated. For human-AI teams to also function in the long term, it should be ensured that the capabilities of both team members remain complementary to each other over time \citep{Spitzer2023}, which could be monitored by employing the measures of our conceptualization. Lastly, attaining CTP also implies that the collaboration mechanism should be consciously designed to improve the realization of inherent and collaborative complementarity potential.

\subsection{Managerial Implications}

Our work also has important implications for managerial decision-makers. In application areas with suitable decision-making tasks, responsible managers should focus on deploying AI systems that enable the realization of CTP through collaboration. If they don’t, their competitors could gain advantages. They could start by collecting data to train AI models that are compatible with teamwork and invest in training their employees in order to enhance their human-AI teams’ inherent complementarity potential.

The conceptualization supports the identification of decision-making use cases suitable for human-AI collaboration by providing the means to measure the complementarity potential and its realized effect. By fostering a deep understanding of the effectiveness of the human-AI team, it enables the targeted improvement of the team’s collaboration. Rather than fearing automation, decision-makers should explore the benefits of working with AI. This is an important perspective, also in the broader societal debate on the role of AI in automation and in the future of work. Our research provides managers with valuable guidance by helping them determine when and how to collaborate with AI to achieve CTP more consistently.

\subsection{Limitations}

Our current research has several limitations that future work needs to address. First, the prevalent use of laboratory-based experiments and vignette methodologies in extant literature on human-AI collaboration in decision-making, including our own, could limit our work’s generalizability and its practical implications. This underscores the need for future studies to use a nuanced approach to prevent contributing to the fragmentation of research in this domain. Furthermore, we focus on developing measurement tools, delineating sources of complementarity, and validating the proposed concepts in behavioral experiments. While our experiments’ controlled settings allow us to derive the insights presented in this work, they do not yet address the wide range of complex human factors, such as motivation \citep{Schunk1995}, engagement \citep{Benz2024,Chandra2022}, self-efficacy \citep{Westphal2024}, or behavior patterns within teams \citep{Schecter2022}, which also contribute to the effectiveness of human-AI teams \citep{Chandra2022}. In this context, many factors in socio-technical systems are highly interconnected and influence the underlying systems’ acceptance and use \citep{Jain2021}. For example, humans could experience emotional reactions towards the AI system or even the task \citep{Han2022,Pfeuffer2019}. In addition, situational factors, such as time pressure, are also likely to affect the human-AI team’s ability to realize complementarity potential \citep{Cao2023,Swaroop2024}. While a broader exploration of these factors is beyond the scope of our work, advancing the field of human-AI collaboration in decision-making further also requires their incorporation as well as finding approaches to manage these factors. We hope that our research will serve as a foundation for studies aiming at expanding knowledge of these mechanisms.

Another limitation is the way in which we measure the counterfactual human decision if the user has not received advice on using AI. In this work, we used a sequential decision-making setup to first measure the human decision and, thereafter, the team decision. However, the sequential nature of the decision-making process could also influence human behavior. Consequently, the timing when the AI’s recommendation is revealed is another critical aspect \citep{Jussupow2021}. Giving the participants the AI's recommendation upfront could lead to cognitive capacity being less invested in the task because the AI has already provided a possible answer \citep{Green2019}.

\subsection{Future Work}

There are several potential areas for future research on human-AI complementarity. The conceptualization could obviously be applied to other domains in the future, but there are also methodological avenues to pursue.

Future work should expand our knowledge about relevant \textit{sources of complementarity}. In this work, we have shown that information and capability asymmetry are the key sources of complementarity, but we have not specified them further. Regarding information asymmetry, we experimentally evaluated unique human contextual information. Investigating unique AI contextual information in future work could also produce interesting insights. Moreover, it would be worthwhile developing criteria regarding the utility of contextual information and the degree of capability asymmetry.

Furthermore, \textit{team settings} could comprise more than two team members, considering multiple AI models or multiple humans. Team design principles could be derived on how to ensure a sufficient degree of complementarity and on how to select and combine human and artificial team members \citep{Hemmer2022}. 

Different \textit{collaboration forms} and \textit{mechanisms} could also be developed and evaluated. While we have focused on humans handling the integration in both experimental studies, the AI might conceivably also undertake the integration. Alternatively, decisions might not be integrated, but task instances might be delegated to a team member deciding on behalf of the team, with the delegation initiated by either the human or the AI \citep{Fugener2021a}. Future work could investigate these collaboration forms in terms of their suitability for specific use cases and application domains.

Finally, although human-AI decision-making is an important application of human-AI collaboration, other types of problems, e.g., creative or generative tasks \citep{Schmidt2023}, may also receive scrutiny with regard to complementarity \citep{Malone2023}.

\section{Conclusion}\label{sec:conclusion_ejis}

So far, human-AI collaboration in decision-making has been primarily concerned with AI systems helping human users. However, since the number of decision tasks that can be automated (i.e., can be solved by the AI alone) is increasing steadily, the focus has shifted to the purposeful design of the collaboration between humans and AI as team members---thereby shaping the future of work with AI. The ultimate objective of these teams must be to achieve complementary team performance (CTP), with the team outperforming each individual team member. The IS community is predestined to drive the development of appropriate theories and to lay the foundation for practical applications. We hope that the conceptual foundation developed in this paper will provide fruitful ground for future research, and that the empirical studies illustrate the validity and potential of the human-AI complementarity paradigm.


\newpage

\bibliographystyle{dcu}

\bibliography{references_new_updated}

@inproceedings{Schmidt2023,
   author = {Corinna Vera Hedwig Schmidt and Michael Guffler and Bastian Kindermann and Tessa Flatten},
   booktitle = {Proceedings of the 44th International Conference on Information Systems},
   pages = {1 - 9},
   title = {{Collaborating with Generative AI: Exploring Algorithm Appreciation in Creative Writing}},
   year = {2023},
}

@article{Turel2023,
   author = {Ofir Turel and Shivam Kalhan},
   issue = {4},
   journal = {MIS Quarterly},
   pages = {1369 - 1394},
   title = {{Prejudiced against the Machine? Implicit Associations and the Transience of Algorithm Aversion.}},
   volume = {47},
   year = {2023},
}

@article{Morrison2024,
   author = {Katelyn Morrison and Philipp Spitzer and Violet Turri and Michelle Feng and Niklas Kühl and Adam Perer},
   issue = {CSCW1},
   journal = {Proceedings of the ACM on Human-Computer Interaction},
   pages = {1 - 39},
   title = {{The Impact of Imperfect XAI on Human-AI Decision-Making}},
   volume = {8},
   year = {2024},
}

@article{Pathirannehelage,
   author = {Savindu Herath Pathirannehelage and Yash Raj Shrestha and Georg von Krogh},
   issue = {Forthcoming},
   journal = {European Journal of Information Systems},
   pages = {1-23},
   publisher = {Taylor & Francis},
   title = {{Design Principles for Artificial Intelligence-Augmented Decision Making: An Action Design Research Study}},
   year = {2024},
}

@article{Collins2021,
   author = {Christopher Collins and Denis Dennehy and Kieran Conboy and Patrick Mikalef},
   journal = {International Journal of Information Management},
   pages = {1-17},
   title = {{Artificial Intelligence in Information Systems Research: A Systematic Literature Review and Research Agenda}},
   volume = {60},
   year = {2021},
}

@article{Foerster,
   author = {Maximilian Förster and Hanna R Broder and Marie C Fahr and Mathias Klier and Lior Fink},
   issue = {Forthcoming},
   journal = {European Journal of Information Systems},
   pages = {1-23},
   publisher = {Taylor & Francis},
   title = {{Tell Me More, Tell Me More: The Impact of Explanations on Learning from Feedback Provided by Artificial Intelligence}},
   year = {2024},
}

@inproceedings{Schoeffer2023,
   author = {Jakob Schoeffer and Johannes Jakubik and Michael Voessing and Niklas Kuehl and Gerhard Satzger},
   booktitle = {Proceedings of the Second International Conference on Hybrid Human-Artificial Intelligence},
   pages = {46-59},
   title = {{On the Interdependence of Reliance Behavior and Accuracy in AI-Assisted Decision-Making}},
   year = {2023},
}

@inproceedings{Schemmer2022,
   author = {Max Schemmer and Patrick Hemmer and Niklas Kühl and Carina Benz and Gerhard Satzger},
   booktitle = {Workshop on Trust and Reliance in AI-Human Teams at the 2022 CHI Conference on Human Factors in Computing Systems},
   pages = {1-10},
   title = {{Should I Follow AI-based Advice? Measuring Appropriate Reliance in Human-AI Decision-Making}},
   year = {2022},
}

@article{Westphal2024,
   author = {Monika Westphal and Patrick Hemmer and Michael Vössing and Max Schemmer and Sebastian Vetter and Gerhard Satzger},
   issue = {Forthcoming},
   journal = {ACM Transactions on Interactive Intelligent Systems},
   note = {Just Accepted},
   pages = {1 - 23},
   title = {{Towards Understanding AI Delegation: The Role of Self-Efficacy and Visual Processing Ability}},
   year = {2024},
}

@article{Arcy,
   author = {John D’Arcy and Ashish Gupta and Monideepa Tarafdar and Ofir Turel},
   issue = {1},
   journal = {Communications of the Association for Information Systems},
   pages = {5},
   title = {{Reflecting on the “Dark Side” of Information Technology Use}},
   volume = {35},
   year = {2014},
}

@article{Parsons2012,
   author = {Jeffrey Parsons and Yair Wand},
   issue = {5},
   journal = {Journal of the Association for Information Systems},
   pages = {245-273},
   title = {{Extending Classification Principles from Information Modeling to Other Disciplines}},
   volume = {14},
   year = {2012},
}

@article{Vericourt,
   author = {Francis de Véricourt and Huseyin Gurkan},
   issue = {Forthcoming},
   journal = {Management Science},
   pages = {1-17},
   title = {{Is Your Machine Better Than You? You May Never Know}},
   year = {2023},
}

@article{Lu2024,
   author = {Tian Lu and Yingjie Zhang},
   issue = {Forthcoming},
   journal = {Information Systems Research},
   title = {{1 + 1 > 2? Information, Humans, and Machines}},
   year = {2024},
}

@article{Rai2019,
   author = {Arun Rai and Panos Constantinides and Saonee Sarker},
   issue = {1},
   journal = {MIS Quarterly},
   pages = {iii-ix},
   title = {{Next Generation Digital Platforms: Toward Human-AI Hybrids}},
   volume = {43},
   year = {2019},
}

@article{Berente2021,
   author = {Nicholas Berente and Bin Gu and Jan Recker and Radhika Santhanam},
   issue = {3},
   journal = {Management Information Systems Quarterly},
   pages = {1433-1450},
   title = {{Special Issue Editor’s Comments: Managing Artificial Intelligence}},
   volume = {45},
   year = {2021},
}

@article{Ancona1992,
   author = {Deborah Gladstein Ancona and David F Caldwell},
   issue = {3},
   journal = {Organization Science},
   month = {10},
   pages = {321-341},
   title = {{Demography and Design: Predictors of New Product Team Performance}},
   volume = {3},
   year = {1992},
}

@article{Gladstein1984,
   author = {Deborah L Gladstein},
   issue = {4},
   journal = {Administrative Science Quarterly},
   pages = {499-517},
   title = {{Groups in Context: A Model of Task Group Effectiveness}},
   volume = {29},
   year = {1984},
}

@article{Hackman1975,
   author = {J.Richard Hackman and Charles G Morris},
   editor = {Leonard B T - Advances in Experimental Social Psychology Berkowitz},
   journal = {Advances in Experimental Social Psychology},
   pages = {45-99},
   title = {{Group Tasks, Group Interaction Process, and Group Performance Effectiveness: A Review and Proposed Integration}},
   volume = {8},
   year = {1975},
}

@article{Spitzer2023,
   author = {Philipp Spitzer and Niklas Kühl and Daniel Heinz and Gerhard Satzger},
   issue = {CSCW2},
   journal = {Proceedings of the ACM on Human-Computer Interaction},
   pages = {1-25},
   title = {{ML-Based Teaching Systems: A Conceptual Framework}},
   volume = {7},
   year = {2023},
}

@article{Power2008,
   author = {Daniel J Power},
   journal = {Handbook on Decision Support Systems 1},
   pages = {121-140},
   title = {{Decision Support Systems: A Historical Overview}},
   year = {2008},
}

@article{Benz2024,
   author = {Carina Benz and Lara Riefle and Gerhard Satzger},
   issue = {1},
   journal = {Communications of the Association for Information Systems},
   pages = {331-359},
   title = {{User Engagement and Beyond: A Conceptual Framework for Engagement in Information Systems Research}},
   volume = {54},
   year = {2024},
}

@article{Kühl_goutier,
   author = {Niklas Kühl and Marc Goutier and Lucas Baier and Clemens Wolff and Dominik Martin},
   journal = {Cognitive Systems Research},
   pages = {78-92},
   title = {{Human vs. Supervised Machine Learning: Who Learns Patterns Faster?}},
   volume = {76},
   year = {2022},
}

@inproceedings{Sarkar2023,
   author = {Bidipta Sarkar and Andy Shih and Dorsa Sadigh},
   booktitle = {Advances in Neural Information Processing Systems},
   pages = {23115-23139},
   title = {{Diverse Conventions for Human-AI Collaboration}},
   volume = {36},
   year = {2023},
}

@article{Inkpen2023,
   author = {Kori Inkpen and Shreya Chappidi and Keri Mallari and Besmira Nushi and Divya Ramesh and Pietro Michelucci and Vani Mandava and Libuše Hannah Vepřek and Gabrielle Quinn},
   issue = {5},
   journal = {ACM Transactions on Computer-Human Interaction},
   pages = {1-29},
   title = {{Advancing Human-AI Complementarity: The Impact of User Expertise and Algorithmic Tuning on Joint Decision Making}},
   volume = {30},
   year = {2023},
}

@inproceedings{Ma2023,
   author = {Shuai Ma and Ying Lei and Xinru Wang and Chengbo Zheng and Chuhan Shi and Ming Yin and Xiaojuan Ma},
   booktitle = {Proceedings of the 2023 CHI Conference on Human Factors in Computing Systems},
   pages = {1-19},
   title = {{Who Should I Trust: AI or Myself? Leveraging Human and AI Correctness Likelihood to Promote Appropriate Trust in AI-Assisted Decision-Making}},
   year = {2023},
}

@article{Dvijotham2023,
   author = {Krishnamurthy (Dj) Dvijotham and Jim Winkens and Melih Barsbey and Sumedh Ghaisas and Robert Stanforth and Nick Pawlowski and Patricia Strachan and Zahra Ahmed and Shekoofeh Azizi and Yoram Bachrach and Laura Culp and Mayank Daswani and Jan Freyberg and Christopher Kelly and Atilla Kiraly and Timo Kohlberger and Scott McKinney and Basil Mustafa and Vivek Natarajan and Krzysztof Geras and Jan Witowski and Zhi Zhen Qin and Jacob Creswell and Shravya Shetty and Marcin Sieniek and Terry Spitz and Greg Corrado and Pushmeet Kohli and Taylan Cemgil and Alan Karthikesalingam},
   issue = {7},
   journal = {Nature Medicine},
   pages = {1814-1820},
   title = {{Enhancing the Reliability and Accuracy of AI-Enabled Diagnosis via Complementarity-Driven Deferral to Clinicians}},
   volume = {29},
   year = {2023},
}

@misc{Cambridge_Dictionary,
   author = {Cambridge Dictionary},
   title = {{English Dictionary}},
   url = {https://dictionary.cambridge.org/us/dictionary/ english/complementarity (Accessed: 2024-07-02)},
   year = {2024},
}

@inproceedings{Brynjolfsson2018,
   author = {Erik Brynjolfsson and Tom Mitchell and Daniel Rock},
   booktitle = {AEA Papers and Proceedings},
   pages = {43-47},
   title = {{What Can Machines Learn and What Does It Mean for Occupations and the Economy?}},
   volume = {108},
   year = {2018},
}

@article{Li2020,
   author = {You Li and Javier Ibanez-Guzman},
   issue = {4},
   journal = {IEEE Signal Processing Magazine},
   pages = {50-61},
   title = {{Lidar for Autonomous Driving: The Principles, Challenges, and Trends for Automotive Lidar and Perception Systems}},
   volume = {37},
   year = {2020},
}

@inproceedings{Swaroop2024,
   author = {Siddharth Swaroop and Zana Buçinca and Krzysztof Z Gajos and Finale Doshi-Velez},
   booktitle = {Proceedings of the 29th International Conference on Intelligent User Interfaces},
   pages = {138–154},
   title = {{Accuracy-Time Tradeoffs in AI-Assisted Decision Making under Time Pressure}},
   year = {2024},
}

@article{Cao2023,
   author = {Shiye Cao and Catalina Gomez and Chien-Ming Huang},
   issue = {CSCW2},
   journal = {Proceedings of the ACM on Human-Computer Interaction},
   pages = {1-26},
   title = {{How Time Pressure in Different Phases of Decision-Making Influences Human-AI Collaboration}},
   volume = {7},
   year = {2023},
}

@article{Pfeuffer2019,
   author = {Nicolas Pfeuffer and Alexander Benlian and Henner Gimpel and Oliver Hinz},
   issue = {4},
   journal = {Business \& Information Systems Engineering},
   pages = {523-533},
   publisher = {Springer},
   title = {{Anthropomorphic Information Systems}},
   volume = {61},
   year = {2019},
}

@article{Han2022,
   author = {Elizabeth Han and Dezhi Yin and Han Zhang},
   issue = {3},
   journal = {Information Systems Research},
   pages = {1296-1311},
   title = {{Bots with Feelings: Should AI Agents Express Positive Emotion in Customer Service?}},
   volume = {34},
   year = {2022},
}

@article{Silver2018,
   author = {David Silver and Thomas Hubert and Julian Schrittwieser and Ioannis Antonoglou and Matthew Lai and Arthur Guez and Marc Lanctot and Laurent Sifre and Dharshan Kumaran and Thore Graepel},
   issue = {6419},
   journal = {Science},
   pages = {1140-1144},
   title = {{A General Reinforcement Learning Algorithm that Masters Chess, Shogi, and Go through Self-play}},
   volume = {362},
   year = {2018},
}

@article{Afshar2022,
   author = {Parnian Afshar and Moezedin Javad Rafiee and Farnoosh Naderkhani and Shahin Heidarian and Nastaran Enshaei and Anastasia Oikonomou and Faranak Babaki Fard and Reut Anconina and Keyvan Farahani and Konstantinos N Plataniotis},
   issue = {1},
   journal = {Scientific Reports},
   pages = {4827-4838},
   title = {{Human-level COVID-19 Diagnosis from Low-dose CT Scans Using a Two-stage Time-distributed Capsule Network}},
   volume = {12},
   year = {2022},
}

@article{Grace2024,
   author = {Katja Grace and Harlan Stewart and Julia Fabienne Sandkühler and Stephen Thomas and Ben Weinstein-Raun and Jan Brauner},
   journal = {arXiv preprint arXiv:2401.02843},
   pages = {1-38},
   title = {{Thousands of AI Authors on the Future of AI}},
   year = {2024},
}

@article{Grace2018,
   author = {Katja Grace and John Salvatier and Allan Dafoe and Baobao Zhang and Owain Evans},
   journal = {Journal of Artificial Intelligence Research},
   pages = {729-754},
   title = {{When Will AI Exceed Human Performance? Evidence from AI Experts}},
   volume = {62},
   year = {2018},
}

@inproceedings{Irvin2019,
   author = {Jeremy Irvin and Pranav Rajpurkar and Michael Ko and Yifan Yu and Silviana Ciurea-Ilcus and Chris Chute and Henrik Marklund and Behzad Haghgoo and Robyn Ball and Katie Shpanskaya and Jayne Seekins and David A Mong and Safwan S Halabi and Jesse K Sandberg and Ricky Jones and David B Larson and Curtis P Langlotz and Bhavik N Patel and Matthew P Lungren and Andrew Y Ng},
   booktitle = {Proceedings of the AAAI Conference on Artificial Intelligence},
   pages = {590-597},
   title = {{CheXpert: A Large Chest Radiograph Dataset with Uncertainty Labels and Expert Comparison}},
   volume = {33},
   year = {2019},
}

@article{Huysman2020,
   author = {Marleen Huysman},
   issue = {4},
   journal = {Journal of Information Technology},
   pages = {307-309},
   title = {{Information Systems Research on Artificial Intelligence and Work: A Commentary on “Robo-Apocalypse Cancelled? Reframing the Automation and Future of Work Debate”}},
   volume = {35},
   year = {2020},
}

@article{Kvaløy2015,
   author = {Ola Kvaløy and Petra Nieken and Anja Schöttner},
   journal = {European Economic Review},
   pages = {188-199},
   title = {{Hidden Benefits of Reward: A Field Experiment on Motivation and Monetary Incentives}},
   volume = {76},
   year = {2015},
}

@article{Sieck2005,
   author = {Winston R Sieck and Hal R Arkes},
   issue = {1},
   journal = {Journal of Behavioral Decision Making},
   pages = {29-53},
   title = {{The Recalcitrance of Overconfidence and its Contribution to Decision Aid Neglect}},
   volume = {18},
   year = {2005},
}

@article{Mahmud2022,
   author = {Hasan Mahmud and A K M Najmul Islam and Syed Ishtiaque Ahmed and Kari Smolander},
   journal = {Technological Forecasting and Social Change},
   pages = {1-26},
   title = {{What Influences Algorithmic Decision-Making? A Systematic Literature Review on Algorithm Aversion}},
   volume = {175},
   year = {2022},
}

@article{Findling2021,
   author = {Charles Findling and Valentin Wyart},
   journal = {Current Opinion in Behavioral Sciences},
   pages = {124-132},
   title = {{Computation Noise in Human Learning and Decision-Making: Origin, Impact, Function}},
   volume = {38},
   year = {2021},
}

@article{Gopnik2012,
   author = {Alison Gopnik and Henry M Wellman},
   issue = {6},
   journal = {Psychological Bulletin},
   pages = {1085-1108},
   title = {{Reconstructing Constructivism: Causal Models, Bayesian Learning Mechanisms, and the Theory Theory.}},
   volume = {138},
   year = {2012},
}

@article{Tenenbaum2011,
   author = {Joshua B Tenenbaum and Charles Kemp and Thomas L Griffiths and Noah D Goodman},
   issue = {6022},
   journal = {Science},
   pages = {1279-1285},
   title = {{How to Grow a Mind: Statistics, Structure, and Abstraction}},
   volume = {331},
   year = {2011},
}

@article{Simons1999,
   author = {Tony Simons and Lisa Hope Pelled and Ken A Smith},
   issue = {6},
   journal = {Academy of Management Journal},
   pages = {662-673},
   title = {{Making Use of Difference: Diversity, Debate, and Decision Comprehensiveness in Top Management Teams}},
   volume = {42},
   year = {1999},
}

@article{Schunk1995,
   author = {Dale H Schunk},
   issn = {1041-3200},
   issue = {2},
   journal = {Journal of Applied Sport Psychology},
   pages = {112-137},
   title = {{Self-Efficacy, Motivation, and Performance}},
   volume = {7},
   year = {1995},
}

@article{Jain2021,
   author = {Hemant Jain and Balaji Padmanabhan and Paul A Pavlou and T S Raghu},
   issue = {3},
   journal = {Information Systems Research},
   pages = {675-687},
   title = {{Editorial for the Special Section on Humans, Algorithms, and Augmented Intelligence: The Future of Work, Organizations, and Society}},
   volume = {32},
   year = {2021},
}

@article{Schecter2022,
   author = {Aaron Schecter and Omid Nohadani and Noshir Contractor},
   issue = {2},
   journal = {Management Information Systems Quarterly},
   pages = {713-738},
   title = {{A Robust Inference Method for Decision-Making in Networks}},
   volume = {46},
   year = {2022},
}

@article{Chandra2022,
   author = {Shalini Chandra and Anuragini Shirish and Shirish C Srivastava},
   issue = {4},
   journal = {Journal of Management Information Systems},
   pages = {969-1005},
   title = {{To Be or Not to Be …Human? Theorizing the Role of Human-Like Competencies in Conversational Artificial Intelligence Agents}},
   volume = {39},
   year = {2022},
}

@article{Dougherty1992,
   author = {Deborah Dougherty},
   issue = {2},
   journal = {Organization Science},
   pages = {179-202},
   title = {{Interpretive Barriers to Successful Product Innovation in Large Firms}},
   volume = {3},
   year = {1992},
}

@article{Cohen1997,
   author = {Susan G Cohen and Diane E Bailey},
   issue = {3},
   journal = {Journal of Management},
   pages = {239-290},
   title = {{What Makes Teams Work: Group Effectiveness Research from the Shop Floor to the Executive Suite}},
   volume = {23},
   year = {1997},
}

@article{Horwitz2005,
   author = {Sujin K Horwitz},
   issue = {2},
   journal = {Human Resource Development Review},
   pages = {219-245},
   title = {{The Compositional Impact of Team Diversity on Performance: Theoretical Considerations}},
   volume = {4},
   year = {2005},
}

@article{Endsley2023,
   author = {Mica R Endsley},
   journal = {Computers in Human Behavior},
   pages = {1-16},
   title = {{Supporting Human-AI Teams:Transparency, Explainability, and Situation Awareness}},
   volume = {140},
   year = {2023},
}

@inproceedings{Madras2018,
   author = {David Madras and Toni Pitassi and Richard Zemel},
   booktitle = {Advances in Neural Information Processing Systems},
   pages = {1-11},
   title = {{Predict Responsibly: Improving Fairness and Accuracy by Learning to Defer}},
   volume = {31},
   year = {2018},
}

@article{Dietvorst2015,
   author = {Berkeley J Dietvorst and Joseph P Simmons and Cade Massey},
   issue = {1},
   journal = {Journal of Experimental Psychology: General},
   pages = {114-126},
   title = {{Algorithm Aversion: People Erroneously Avoid Algorithms after Seeing Them Err.}},
   volume = {144},
   year = {2015},
}

@article{Faul2007,
   author = {Franz Faul and Edgar Erdfelder and Albert-Georg Lang and Axel Buchner},
   issue = {2},
   journal = {Behavior Research Methods},
   pages = {175-191},
   title = {{G*Power 3: A Flexible Statistical Power Analysis Program for the Social, Behavioral, and Biomedical Sciences}},
   volume = {39},
   year = {2007},
}

@article{Bauer2023a,
   author = {Kevin Bauer and Moritz von Zahn and Oliver Hinz},
   issue = {4},
   journal = {Information Systems Research},
   pages = {1582-1602},
   title = {{Expl(AI)ned: The Impact of Explainable Artificial Intelligence on Users’ Information Processing}},
   volume = {34},
   year = {2023},
}

@inproceedings{Huang2017,
   author = {Gao Huang and Zhuang Liu and Laurens Van Der Maaten and Kilian Q Weinberger},
   booktitle = {Proceedings of the Conference on Computer Vision and Pattern Recognition},
   pages = {4700-4708},
   title = {{Densely Connected Convolutional Networks}},
   year = {2017},
}

@article{Russakovsky2015,
   author = {Olga Russakovsky and Jia Deng and Hao Su and Jonathan Krause and Sanjeev Satheesh and Sean Ma and Zhiheng Huang and Andrej Karpathy and Aditya Khosla and Michael Bernstein and Alexander C Berg and Li Fei-Fei},
   issue = {3},
   journal = {International Journal of Computer Vision},
   pages = {211-252},
   title = {{ImageNet Large Scale Visual Recognition Challenge}},
   volume = {115},
   year = {2015},
}

@article{Geirhos2020,
   author = {Robert Geirhos and Jörn-Henrik Jacobsen and Claudio Michaelis and Richard Zemel and Wieland Brendel and Matthias Bethge and Felix A Wichmann},
   issue = {11},
   journal = {Nature Machine Intelligence},
   pages = {665-673},
   title = {{Shortcut Learning in Deep Neural Networks}},
   volume = {2},
   year = {2020},
}

@article{Kordzadeh2022,
   author = {Nima Kordzadeh and Maryam Ghasemaghaei},
   issue = {3},
   journal = {European Journal of Information Systems},
   pages = {388-409},
   title = {{Algorithmic Bias: Review, Synthesis, and Future Research Directions}},
   volume = {31},
   year = {2022},
}

@article{Vassilakopoulou2023,
   author = {Polyxeni Vassilakopoulou and Arve Haug and Leif Martin Salvesen and Ilias O Pappas},
   issue = {1},
   journal = {European Journal of Information Systems},
   pages = {10-22},
   title = {{Developing Human/AI Interactions for Chat-Based Customer Services: Lessons Learned from the Norwegian Government}},
   volume = {32},
   year = {2023},
}

@article{Gronsund2020,
   author = {Tor Grønsund and Margunn Aanestad},
   issue = {2},
   journal = {The Journal of Strategic Information Systems},
   pages = {1-16},
   title = {{Augmenting the Algorithm: Emerging Human-in-the-loop Work Configurations}},
   volume = {29},
   year = {2020},
}

@article{Rinta-Kahila2022,
   author = {Tapani Rinta-Kahila and Ida Someh and Nicole Gillespie and Marta Indulska and Shirley Gregor},
   issue = {3},
   journal = {European Journal of Information Systems},
   pages = {313-338},
   title = {{Algorithmic Decision-Making and System Destructiveness: A Case of Automatic Debt Recovery}},
   volume = {31},
   year = {2022},
}

@article{Kleinberg2018,
   author = {Jon Kleinberg and Himabindu Lakkaraju and Jure Leskovec and Jens Ludwig and Sendhil Mullainathan},
   issue = {1},
   journal = {Quarterly Journal of Economics},
   pages = {237-293},
   title = {{Human Decisions and Machine Predictions}},
   volume = {133},
   year = {2018},
}

@article{Mikalef2021,
   author = {Patrick Mikalef and Manjul Gupta},
   issue = {3},
   journal = {Information \& Management},
   pages = {1-20},
   title = {{Artificial Intelligence Capability: Conceptualization, Measurement Calibration, and Empirical Study on its Impact on Organizational Creativity and Firm Performance}},
   volume = {58},
   year = {2021},
}

@article{Mikalef2022,
   author = {Patrick Mikalef and Kieran Conboy and Jenny Eriksson Lundström and Aleš Popovič},
   issue = {3},
   journal = {European Journal of Information Systems},
   pages = {257-268},
   title = {{Thinking Responsibly about Responsible AI and ‘the Dark Side’ of AI}},
   volume = {31},
   year = {2022},
}

@article{Green2019,
   author = {Ben Green and Yiling Chen},
   issue = {CSCW},
   journal = {Proceedings of the ACM on Human-Computer Interaction},
   pages = {1–24},
   title = {{The Principles and Limits of Algorithm-in-the-Loop Decision Making}},
   volume = {3},
   year = {2019},
}

@inproceedings{Bubeck2023,
   author = {Sébastien Bubeck and Varun Chandrasekaran and Ronen Eldan and Johannes Gehrke and Eric Horvitz and Ece Kamar and Peter Lee and Yin Tat Lee and Yuanzhi Li and Scott Lundberg},
   booktitle = {arXiv preprint arXiv:2303.12712},
   pages = {1-155},
   title = {{Sparks of Artificial General Intelligence: Early Experiments with GPT-4}},
   year = {2023},
}

@article{Longoni2019,
   author = {Chiara Longoni and Andrea Bonezzi and Carey K Morewedge},
   issue = {4},
   journal = {Journal of Consumer Research},
   pages = {629-650},
   title = {{Resistance to Medical Artificial Intelligence}},
   volume = {46},
   year = {2019},
}

@article{Rousseeuw1993,
   author = {Peter J Rousseeuw and Christophe Croux},
   issue = {424},
   journal = {Journal of the American Statistical Association},
   pages = {1273-1283},
   title = {{Alternatives to the Median Absolute Deviation}},
   volume = {88},
   year = {1993},
}

@article{Leys2013,
   author = {Christophe Leys and Christophe Ley and Olivier Klein and Philippe Bernard and Laurent Licata},
   issue = {4},
   journal = {Journal of Experimental Social Psychology},
   pages = {764-766},
   title = {{Detecting Outliers: Do Not Use Standard Deviation around the Mean, Use Absolute Deviation around the Median}},
   volume = {49},
   year = {2013},
}

@article{Grootswagers2020,
   author = {Tijl Grootswagers},
   issue = {6},
   journal = {Behavior Research Methods},
   pages = {2283-2286},
   title = {{A Primer on Running Human Behavioural Experiments Online}},
   volume = {52},
   year = {2020},
}

@inproceedings{Wilder2020,
   author = {Bryan Wilder and Eric Horvitz and Ece Kamar},
   booktitle = {Proceedings of the Twenty-Ninth International Joint Conference on Artificial Intelligence},
   pages = {1526-1533},
   title = {{Learning to Complement Humans}},
   year = {2020},
}

@inproceedings{Hemmer2022,
   author = {Patrick Hemmer and Sebastian Schellhammer and Michael Vössing and Johannes Jakubik and Gerhard Satzger},
   booktitle = {Proceedings of the Thirty-First International Joint Conference on Artificial Intelligence},
   pages = {2478-2484},
   title = {{Forming Effective Human-AI Teams: Building Machine Learning Models that Complement the Capabilities of Multiple Experts}},
   year = {2022},
}

@inproceedings{Ross2023,
   author = {Steven I Ross and Fernando Martinez and Stephanie Houde and Michael Muller and Justin D Weisz},
   booktitle = {Proceedings of the 28th International Conference on Intelligent User Interfaces},
   pages = {491-514},
   title = {{The Programmer’s Assistant: Conversational Interaction with a Large Language Model for Software Development}},
   year = {2023},
}

@article{Turiel2020,
   author = {J D Turiel and T Aste},
   issue = {6},
   journal = {Royal Society Open Science},
   pages = {1-17},
   title = {{Peer-to-Peer Loan Acceptance and Default Prediction with Artificial Intelligence}},
   volume = {7},
   year = {2020},
}

@article{Reverberi2022,
   author = {Carlo Reverberi and Tommaso Rigon and Aldo Solari and Cesare Hassan and Paolo Cherubini and Andrea Cherubini},
   issue = {1},
   journal = {Scientific Reports},
   pages = {1-10},
   title = {{Experimental Evidence of Effective Human–AI Collaboration in Medical Decision-Making}},
   volume = {12},
   year = {2022},
}

@article{LeCun2015,
   author = {Yann LeCun and Yoshua Bengio and Geoffrey Hinton},
   issue = {7553},
   journal = {Nature},
   pages = {436-444},
   title = {{Deep Learning}},
   volume = {521},
   year = {2015},
}

@inproceedings{He2023,
   author = {Gaole He and Lucie Kuiper and Ujwal Gadiraju},
   booktitle = {Proceedings of the 2023 CHI Conference on Human Factors in Computing Systems},
   pages = {1-18},
   title = {{Knowing About Knowing: An Illusion of Human Competence Can Hinder Appropriate Reliance on AI Systems}},
   year = {2023},
}

@article{Terveen1995,
   author = {Loren G Terveen},
   issue = {2},
   journal = {Knowledge-Based Systems},
   pages = {67-81},
   title = {{Overview of Human-Computer Collaboration}},
   volume = {8},
   year = {1995},
}

@inproceedings{Geirhos2021,
   author = {Robert Geirhos and Kantharaju Narayanappa and Benjamin Mitzkus and Tizian Thieringer and Matthias Bethge and Felix A Wichmann and Wieland Brendel},
   booktitle = {Advances in Neural Information Processing Systems},
   pages = {23885-23899},
   title = {{Partial Success in Closing the Gap between Human and Machine Vision}},
   volume = {34},
   year = {2021},
}

@article{Goldenberg2019,
   author = {S Larry Goldenberg and Guy Nir and Septimiu E Salcudean},
   issue = {7},
   journal = {Nature Reviews Urology},
   pages = {391-403},
   title = {{A New Era: Artificial Intelligence and Machine Learning in Prostate Cancer}},
   volume = {16},
   year = {2019},
}

@inproceedings{Lai2021,
   author = {Vivian Lai and Chacha Chen and Alison Smith-Renner and Q Vera Liao and Chenhao Tan},
   booktitle = {Proceedings of the 2023 ACM Conference on Fairness, Accountability, and Transparency},
   pages = {1369–1385},
   title = {{Towards a Science of Human-AI Decision Making: An Overview of Design Space in Empirical Human-Subject Studies}},
   year = {2023},
}

@article{Vossing2022,
   author = {Michael Vössing and Niklas Kühl and Matteo Lind and Gerhard Satzger},
   issue = {3},
   journal = {Information Systems Frontiers},
   pages = {877-895},
   title = {{Designing Transparency for Effective Human-AI Collaboration}},
   volume = {24},
   year = {2022},
}

@article{Kuhl2022,
   author = {Niklas Kühl and Max Schemmer and Marc Goutier and Gerhard Satzger},
   issue = {4},
   journal = {Electronic Markets},
   pages = {2235–2244},
   title = {{Artificial Intelligence and Machine Learning}},
   volume = {32},
   year = {2022},
}

@article{Steyvers2022,
   author = {Mark Steyvers and Heliodoro Tejeda and Gavin Kerrigan and Padhraic Smyth},
   issue = {11},
   journal = {Proceedings of the National Academy of Sciences},
   pages = {1-7},
   title = {{Bayesian Modeling of Human–AI Complementarity}},
   volume = {119},
   year = {2022},
}

@article{Malone2023,
   author = {Thomas Malone and Michelle Vaccaro and Andres Campero and Jaeyoon Song and Haoran Wen and Abdullah Almaatouq},
   journal = {Preprint},
   pages = {1-13},
   title = {{A Test for Evaluating Performance in Human-AI Systems}},
   year = {2023},
}

@article{Breiman2001,
   author = {Leo Breiman},
   issue = {1},
   journal = {Machine Learning},
   pages = {5-32},
   title = {{Random Forests}},
   volume = {45},
   year = {2001},
}

@misc{Kaggle2019,
   author = {Kaggle},
   title = {{House Prices and Images - SoCal}},
   url = {https://www.kaggle.com/ted8080/house-prices-and-images-socal (Accessed: 2021-06-16)},
   year = {2019},
}

@article{Jussupow2020,
   author = {Ekaterina Jussupow and Izak Benbasat and Armin Heinzl},
   journal = {Proceedings of the 28th European Conference on Information Systems},
   pages = {1-16},
   title = {{Why Are We Averse Towards Algorithms? A Comprehensive Literature Review on Algorithm Aversion}},
   year = {2020},
}

@inproceedings{Donahue2022,
   author = {Kate Donahue and Alexandra Chouldechova and Krishnaram Kenthapadi},
   booktitle = {Proceedings of the 2022 Conference on Fairness, Accountability, and Transparency},
   pages = {1639-1656},
   title = {{Human-Algorithm Collaboration: Achieving Complementarity and Avoiding Unfairness}},
   year = {2022},
}

@inproceedings{Rastogi2023,
   author = {Charvi Rastogi and Liu Leqi and Kenneth Holstein and Hoda Heidari},
   booktitle = {Proceedings of the AAAI Conference on Human Computation and Crowdsourcing},
   pages = {127-139},
   title = {{A Taxonomy of Human and ML Strengths in Decision-Making to Investigate Human-ML Complementarity}},
   year = {2023},
}

@article{Jussupow2021,
   author = {Ekaterina Jussupow and Kai Spohrer and Armin Heinzl and Joshua Gawlitza},
   issue = {3},
   journal = {Information Systems Research},
   pages = {713-735},
   title = {{Augmenting Medical Diagnosis Decisions? An Investigation into Physicians’ Decision-Making Process with Artificial Intelligence}},
   volume = {32},
   year = {2021},
}

@article{Hillman2019,
   author = {Noel L Hillman},
   issue = {1},
   journal = {Judges' Journal},
   pages = {36-39},
   title = {{The Use of Artificial Intelligence in Gauging the Risk of Recidivism}},
   volume = {58},
   year = {2019},
}

@article{McKinney2020,
   author = {Scott Mayer McKinney and Marcin Sieniek and Varun Godbole and Jonathan Godwin and Natasha Antropova and Hutan Ashrafian and Trevor Back and Mary Chesus and Greg S Corrado and Ara Darzi},
   issue = {7788},
   journal = {Nature},
   pages = {89-94},
   title = {{International Evaluation of an AI System for Breast Cancer Screening}},
   volume = {577},
   year = {2020},
}

@article{Sanders1995,
   author = {Nada R Sanders and Larry P Ritzman},
   issue = {4},
   journal = {Journal of Operations Management},
   pages = {311-321},
   title = {{Bringing Judgment into Combination Forecasts}},
   volume = {13},
   year = {1995},
}

@article{Fugener2021a,
   author = {Andreas Fügener and Jörn Grahl and Alok Gupta and Wolfgang Ketter},
   issue = {2},
   journal = {Information Systems Research},
   pages = {678-696},
   title = {{Cognitive Challenges in Human–Artificial Intelligence Collaboration: Investigating the Path Toward Productive Delegation}},
   volume = {33},
   year = {2022},
}

@inproceedings{Mozannar2020,
   author = {Hussein Mozannar and David Sontag},
   booktitle = {Proceedings of the International Conference on Machine Learning},
   pages = {7076-7087},
   title = {{Consistent Estimators for Learning to Defer to an Expert}},
   year = {2020},
}

@article{Sanders2001,
   author = {Nada R Sanders and Larry P Ritzman},
   journal = {Principles of Forecasting},
   pages = {405-416},
   title = {{Judgmental Adjustment of Statistical Forecasts}},
   year = {2001},
}

@article{Sanders1991,
   author = {Nada R Sanders and Larry P Ritzman},
   issue = {6},
   journal = {International Journal of Operations \& Production Management},
   pages = {27-37},
   title = {{On Knowing When to Switch from Quantitative to Judgemental Forecasts}},
   volume = {11},
   year = {1991},
}

@inproceedings{Gerber2020,
   author = {Alina Gerber and Patrick Derckx and Daniel A. Döppner and Detlef Schoder},
   booktitle = {Proceedings of the 53rd Hawaii International Conference on System Sciences},
   pages = {289-298},
   title = {{Conceptualization of the Human-Machine Symbiosis – A Literature Review}},
   volume = {3},
   year = {2020},
}

@article{Licklider1960,
   author = {Joseph C R Licklider},
   issue = {1},
   journal = {IRE Transactions on Human Factors in Electronics},
   pages = {4-11},
   title = {{Man-Computer Symbiosis}},
   volume = {HFE-1},
   year = {1960},
}

@article{Zhou2021,
   author = {Lina Zhou and Souren Paul and Haluk Demirkan and Lingyao Yuan and Jim Spohrer and Michelle Zhou and Julie Basu},
   issue = {2},
   journal = {AIS Transactions on Human-Computer Interaction},
   pages = {243-264},
   title = {{Intelligence Augmentation: Towards Building Human-Machine Symbiotic Relationship}},
   volume = {13},
   year = {2021},
}

@inproceedings{Zhang2022,
   author = {Qiaoning Zhang and Matthew L Lee and Scott Carter},
   booktitle = {Proceedings of the 2022 CHI Conference on Human Factors in Computing Systems},
   pages = {1-28},
   title = {{You Complete Me: Human-AI Teams and Complementary Expertise}},
   year = {2022},
}

@inproceedings{Ribeiro2018,
   author = {Marco Tulio Ribeiro and Sameer Singh and Carlos Guestrin},
   issue = {1},
   booktitle = {Proceedings of the AAAI Conference on Artificial Intelligence},
   note = {select best classifier},
   pages = {1527-1535},
   title = {{Anchors : High-Precision Model-Agnostic Explanations}},
   volume = {32},
   year = {2018},
}

@inproceedings{Ribeiro2016,
   author = {Marco Tulio Ribeiro and Sameer Singh and Carlos Guestrin},
   booktitle = {Proceedings of the 22nd ACM SIGKDD International Conference on Knowledge Discovery and Data Mining},
   pages = {1135-1144},
   title = {{"Why Should I Trust You?" Explaining the Predictions of Any Classifier}},
   year = {2016},
}

@article{Adadi2018,
   author = {Amina Adadi and Mohammed Berrada},
   journal = {IEEE Access},
   pages = {52138-52160},
   title = {{Peeking Inside the Black-Box: A Survey on Explainable Artificial Intelligence (XAI)}},
   volume = {6},
   year = {2018},
}

@inproceedings{Kunkel2019,
   author = {Johannes Kunkel and Tim Donkers and Lisa Michael and Catalin-Mihai Barbu and Jürgen Ziegler},
   booktitle = {Proceedings of the 2019 CHI Conference on Human Factors in Computing Systems},
   pages = {1-12},
   title = {{Let Me Explain: Impact of Personal and Impersonal Explanations on Trust in Recommender Systems}},
   year = {2019},
}

@inproceedings{Yu2019,
   author = {Kun Yu and Shlomo Berkovsky and Ronnie Taib and Jianlong Zhou and Fang Chen},
   booktitle = {Proceedings of the 24th International Conference on Intelligent User Interfaces},
   pages = {460-468},
   title = {{Do I Trust My Machine Teammate? An Investigation from Perception to Decision}},
   year = {2019},
}

@article{VanderWaa2021,
   author = {Jasper van der Waa and Elisabeth Nieuwburg and Anita Cremers and Mark Neerincx},
   journal = {Artificial Intelligence},
   pages = {1-19},
   title = {{Evaluating XAI: A Comparison of Rule-Based and Example-Based Explanations}},
   volume = {291},
   year = {2021},
}

@inproceedings{Alufaisan2021,
   author = {Yasmeen Alufaisan and Laura R Marusich and Jonathan Z Bakdash and Yan Zhou and Murat Kantarcioglu},
   issue = {8},
   booktitle = {Proceedings of the AAAI Conference on Artificial Intelligence},
   pages = {6618-6626},
   title = {{Does Explainable Artificial Intelligence Improve Human Decision-Making?}},
   volume = {35},
   year = {2021},
}

@inproceedings{Bucinca2020,
   author = {Zana Buçinca and Phoebe Lin and Krzysztof Z. Gajos and Elena L. Glassman},
   booktitle = {Proceedings of the 25th International Conference on Intelligent User Interfaces},
   pages = {454-464},
   title = {{Proxy Tasks and Subjective Measures Can Be Misleading in Evaluating Explainable AI Systems}},
   year = {2020},
}

@inproceedings{Carton2020,
   author = {Samuel Carton and Qiaozhu Mei and Paul Resnick},
   issue = {1},
   booktitle = {Proceedings of the International AAAI Conference on Web and Social Media},
   pages = {95-106},
   title = {{Feature-Based Explanations Don't Help People Detect Misclassifications of Online Toxicity}},
   volume = {14},
   year = {2020},
}

@article{Ibrahim2021,
   author = {Rouba Ibrahim and Song-Hee Kim and Jordan Tong},
   issue = {4},
   journal = {Management Science},
   pages = {2314-2325},
   title = {{Eliciting Human Judgment for Prediction Algorithms}},
   volume = {67},
   year = {2021},
}

@article{Fugener2021,
   author = {Andreas Fügener and Jörn Grahl and Alok Gupta and Wolfgang Ketter},
   issue = {3},
   journal = {Management Information Systems Quarterly},
   pages = {1527-1556},
   title = {{Will Humans-in-the-Loop Become Borgs? Merits and Pitfalls of Working with AI}},
   volume = {45},
   year = {2021},
}

@inproceedings{Zhang2020,
   author = {Yunfeng Zhang and Q Vera Liao and Rachel K E Bellamy},
   booktitle = {Proceedings of the 2020 Conference on Fairness, Accountability, and Transparency},
   pages = {295-305},
   title = {{Effect of Confidence and Explanation on Accuracy and Trust Calibration in AI-Assisted Decision Making}},
   year = {2020},
}

@inproceedings{Lai2020,
   author = {Vivian Lai and Han Liu and Chenhao Tan},
   booktitle = {Proceedings of the 2020 CHI Conference on Human Factors in Computing Systems},
   pages = {1-13},
   title = {{Why Is 'Chicago' Deceptive? Towards Building Model-Driven Tutorials for Humans}},
   year = {2020},
}

@article{Liu2021,
   author = {Han Liu and Vivian Lai and Chenhao Tan},
   issue = {CSCW2},
   journal = {Proceedings of the ACM on Human-Computer Interaction},
   pages = {1-45},
   title = {{Understanding the Effect of Out-of-Distribution Examples and Interactive Explanations on Human-AI Decision Making}},
   volume = {5},
   year = {2021},
}

@inproceedings{Hemmer2021,
   author = {Patrick Hemmer and Max Schemmer and Michael Vössing and Niklas Kühl},
   booktitle = {Proceedings of the 25th Pacific Asia Conference on Information Systems},
   pages = {1-14},
   title = {{Human-AI Complementarity in Hybrid Intelligence Systems: A Structured Literature Review}},
   year = {2021},
}

@inproceedings{Bansal2021,
   author = {Gagan Bansal and Tongshuang Wu and Joyce Zhou and Raymond Fok and Besmira Nushi and Ece Kamar and Marco Tulio Ribeiro and Daniel Weld},
   booktitle = {Proceedings of the 2021 CHI Conference on Human Factors in Computing Systems},
   pages = {1-16},
   title = {{Does the Whole Exceed its Parts? The Effect of AI Explanations on Complementary Team Performance}},
   year = {2021},
}

@article{Lake2017,
   author = {Brenden M Lake and Tomer D Ullman and Joshua B Tenenbaum and Samuel J Gershman},
   journal = {Behavioral and Brain Sciences},
   pages = {1-72},
   title = {{Building Machines that Learn and Think like People}},
   volume = {40},
   year = {2017},
}

@article{Dellermann2019,
   author = {Dominik Dellermann and Philipp Ebel and Matthias Söllner and Jan Marco Leimeister},
   issue = {5},
   journal = {Business \& Information Systems Engineering},
   pages = {637-643},
   title = {{Hybrid Intelligence}},
   volume = {61},
   year = {2019},
}

@article{Seeber2020,
   author = {Isabella Seeber and Eva Bittner and Robert O Briggs and Triparna De Vreede and Gert-Jan De Vreede and Aaron Elkins and Ronald Maier and Alexander B Merz and Sarah Oeste-Reiß and Nils Randrup},
   issue = {2},
   journal = {Information \& Management},
   pages = {1-22},
   title = {{Machines as Teammates: A Research Agenda on AI in Team Collaboration}},
   volume = {57},
   year = {2020},
}

@article{Stauder2021,
   author = {Maximilian Stauder and Niklas Kühl},
   issue = {3},
   journal = {Flexible Services and Manufacturing Journal},
   pages = {709–747},
   title = {{AI for In-Line Vehicle Sequence Controlling: Development and Evaluation of an Adaptive Machine Learning Artifact to Predict Sequence Deviations in a Mixed-Model Production Line}},
   volume = {34},
   year = {2022},
}

@inproceedings{Day2018,
   author = {Min-Yuh Day and Tun-Kung Cheng and Jheng-Gang Li},
   booktitle = {2018 IEEE/ACM International Conference on Advances in Social Networks Analysis and Mining},
   pages = {1027-1031},
   title = {{AI Robo-Advisor with Big Data Analytics for Financial Services}},
   year = {2018},
}

@inproceedings{Mallari2020,
   author = {Keri Mallari and Kori Inkpen and Paul Johns and Sarah Tan and Divya Ramesh and Ece Kamar},
   booktitle = {Proceedings of the 2020 CHI Conference on Human Factors in Computing Systems},
   pages = {1-13},
   title = {{Do I Look like a Criminal? Examining how Race Presentation Impacts Human Judgement of Recidivism}},
   year = {2020},
}




\end{document}